\documentclass[aps,prd,showpacs,nofootinbib,floatfix,amsmath,amssymb]{revtex4-2}

\usepackage{graphicx,amsfonts}
\usepackage{epsfig}
\usepackage{bm}
\usepackage{siunitx}
\usepackage[hidelinks]{hyperref}
\usepackage{float}
\usepackage[export]{adjustbox}
\usepackage{subcaption}
\usepackage{array,booktabs}
\usepackage{bbold}
\DeclareUnicodeCharacter{2212}{-}

\begin{document}
	
\title{Constraints on exotic particle masses from flavor violating charged lepton decays and the role of interference}
	
\author{Mario W. Barela}\email{mario.barela@unesp.br}
\affiliation{
Instituto de F\'isica Te\'orica, Universidade Estadual Paulista, \\
R. Dr. Bento Teobaldo Ferraz 271, Barra Funda\\ S\~ao Paulo - SP, 01140-070,
Brazil.}

\author{J. Monta\~no-Dom\'inguez}\email{jmontano@conacyt.mx}
\affiliation{
Conacyt--Facultad de Ciencias F\'isico Matem\'aticas, Universidad Michoacana de San Nicol\'as de Hidalgo,
Av. Francisco J. M\'ugica s/n, 58060, Morelia, Michoac\'an, M\'exico.
}

\begin{abstract}
Together with any Beyond the Standard Model ultraviolet complete, renormalizable model, one or several exotic particles are usually hypothesized. $C\!P$-even neutral and doubly-charged scalars are common, well known examples which can contribute to the seven 3-body Charged Lepton Flavor Violating decays. The experimental bounds on each Branching Ratio within this set of processes provide a good test for new physics that can induce powerful constraints on relevant parameter spaces. This is specially true for a third species of particle which, unlike the previous two, is a rare feature of renormalizable models: a doubly charged vector bilepton. We show how these purely leptonic bounds can indeed induce relevant exclusion regions for the corresponding particles masses, stronger than what have been considered in the literature for the alternative flavor conserving case, and examine how interference effects can influence these regions in a non-trivial way.
\end{abstract}

\maketitle

\section{Introduction}
\label{sec:intro}

Despite its phenomenal success in justifying and describing some of the most complex experimental results, including high precision data and the confirmed Higgs boson prediction, the Standard Model (SM) of elementary particles and fundamental interactions presents several unavoidable problems. These include the inability of accounting for urgent physical issues, such as Neutrino Masses, Dark Matter and Matter-Antimatter Asymmetry. Additionally, it contains problematic aspects which may be interpreted as matters of non-optimal construction, such as the Hierarchy Problem or the arbitrariness in the number of families. From this, emerged the necessity of Beyond the Standard Model (BSM) physics, whose classes of most elegant tentative instances contain Grand Unified Theories (GUT) and Supersymmetry. Once these types of theories have, at least, not been encouraged at the LHC scales, models with, in principle, lower characteristic scales, conceptually similar to the SM, with a non-simple Gauge Group, have began to be examined.

Independently of the species of extension or alternative model, the new constructions generally imply the introduction of new degrees of freedom, usually not immediately desirable by themselves. Among the examples of new physics generated together with exotic particles is Charged Lepton Flavor Violation (CLFV). CLFV signatures are forbidden at tree level in the context of the SM, and hence are extremely useful in the quest of testing alternative hypotheses. Specifically, they can occur, in the SM, at the one-loop level and are, naturally, extremely suppressed: it is predicted that the main muon and tau branching ratios, $\mu\to 3e$, $\tau\to 3e$ and $\tau\to 3\mu$, for example, happen with order of magnitude $<10^{-50}$ \cite{Petcov:1976ff,deGouvea:2013zba,Heeck:2016xwg,Hernandez-Tome:2018fbq}. To date, there is no concrete direct sign of CLFV, and the status of this type of search is of improving sensitivity \cite{Bernstein:2013hba,Mori:2014aqa,Calibbi:2017uvl,MEGII:2018kmf,Kraetzschmar:2021rly,
Guzzi:2020fmg,Hayasaka:2010np}.


Besides being efficient because of virtually nonexistent SM background, CLFV is also a great prospect for BSM searches because it may be contained in most kinds of new physics models, such as Supersymmetry \cite{Romao:1984pn,Borzumati:1986qx,Martin:1997ns,Altmannshofer:2009ne}, 2HDM \cite{Davidson:2010xv,Dery:2013aba,Kopp:2014rva} and 3-3-1 models \cite{Cabarcas:2012uf, Liu:1993gy,Machado:2016jzb,Barela:2019pmo}. 

An interesting specific set of CLFV channels is that of the already mentioned 3-body lepton decays, \textit{i.e.}, the decays of muons and taus into respective combinations of 3 lighter leptons. These processes enjoy the operational benefits of being purely leptonic: they are free of a number of complications present in the analysis of LHC processes: in particular, hadron physics, which poses both computational \textit{and} theoretical difficulties. For instance, it is hard to effect truly general and model-independent analysis on a number of BSM LHC processes because of highly model-dependent hadronization structures.

The objective of this work is to investigate whether the simultaneous application of the bounds on the Branching Ratios (BR) of 3-body lepton decays is enough to achieve useful results -- or isolate the sector of theory space where said results are valuable -- regarding the masses of three species of exotic particles: an exotic flavor violating neutral scalar $s$, a doubly-charged scalar bilepton $Y^{\pm \pm}$ and a doubly-charged vector bilepton $U^{\pm \pm}$. Most important in our discussion is the vector bilepton, a rare particle present most notably in 3-3-1 models \cite{Pisano:1992bxx,Frampton:1992wt,Dias:2006ns} and $SU(15)$ GUT \cite{Frampton:1989fu,Frampton:1990hz}, which entails unitary mixing. Nevertheless, special attention will be paid on the occurring interferences when two distinct bosons are mediating the decays, to observe that such effects can impact in a meaningful way bounds on new BSM degrees of freedom.

The paper is organized as follows. In Section \ref{sec:effint} relevant discussions concerning the model-independent parametrization of the relevant interactions are carried, and the corresponding effective Lagrangians presented. In Sec. \ref{sec:compmet} the charged LFV decay amplitudes contributing to a generic $\ell^+_l(p)\to \ell^+_i(k_1)\ell^+_j(k_2)\ell^-_k(k_3)$ are shown. The Sec.~\ref{sec:res} is devoted to show the results and discussing its consequences. Finally, our conclusions are addressed in Sec. \ref{sec:conc}. In the Appendix \ref{app:amps} the non-trivial attainment of the amplitudes from Lagrangians that contain explicit charge conjugation is described, as well the corresponding Feynman rules presented. In the Appendix \ref{app:sol}, some numerical solutions are explicitly written.

\section{Objectives and effective interactions}
\label{sec:effint}

This work's main goal is to draw relevant exclusion contours on the masses of three different species of exotic particles, constraining theory space by taking advantage of the simultaneous data of 3-body lepton decay. The examined particles are the ones which can contribute to the relevant processes: doubly-charged vector bileptons $U^{\pm \pm}$; doubly-charged scalars $Y^{\pm \pm}$; and flavor violation mediating neutral scalars $s$.

In order to take into account and gain insight on interesting interference effects \cite{Barela:2021pzv} and because it can be expected that there is more than just one exotic particle not contemplated by the SM, we consider two particles at a time.

Besides the aforementioned particles, a neutral vector boson $Z^\prime$ could also contribute to the significant processes. This, however, can only happen in non-democratic underlying models, where distinct lepton families constitute different representations of the gauge group, otherwise the a priori diagonal kinetic terms result in a mixing matrix of the form $\mathcal{O}_{Z^\prime}=V_L^\dagger V_L=\mathbb{1}$. Because of this, since we avoid focusing on specific models and, furthermore, non-democratic leptonic sectors being rare, we overlook the possible role of an exotic neutral vector boson.

A substantial aspect of the challenge resumes itself to a correct parametrization of the effective exotic interactions. Consider first the doubly-charged vector bilepton $U^{\pm \pm}$. The possible electric charge and handedness structure of the spinor chain limits the form of the most general  $U\ell\ell$ interaction Lagrangian to be
\begin{equation}
\mathcal{L}_{U\ell\ell}=\sum_a g_U \overline{\ell_{a\phantom{c}L}^{\prime c}}\gamma^\mu \ell'_{aL}\thinspace U_\mu^{++}
+ g_U\overline{\ell'_{aL}}\gamma^\mu \ell^{\prime c}_{a\phantom{c}L}\thinspace U_\mu^{--},
\end{equation}
which is diagonal on lepton symmetry eigenstates, the primed fields, because it comes from a minimally coupled kinetic term and we only consider lepton universal models, as discussed above, and where the second term is merely the hermitian conjugate of the first one. Notice that unlike interactions which conserve fermion number, any diagonal vertex from the sum exhausts the degrees of freedom of a single fermion. Hence, we may consider only Lagrangian above, without writing a hand-mirrored term $g^\prime_U \overline{\ell_{a\phantom{c}R}^{\prime c}}\gamma^\mu \ell'_{aR} U_\mu^{++}$  that would involve the same fields. In fact, $g^\prime_U \overline{\ell_{a\phantom{c}R}^{\prime c}}\gamma^\mu \ell'_{aR} U_\mu^{++} = -g^\prime_U \overline{\ell_{a\phantom{c}L}^{\prime c}}\gamma^\mu \ell'_{aL} U_\mu^{++}$, so that adding the second term would amount to a mere redefinition of the coupling.

The fermions are rotated to their mass eigenstates through the bi-unitary transformation $\ell'_{L(R)} \equiv V_{L(R)} \ell_{L(R)}$, from which follows (taking left-handed fields as representative)
\begin{equation}
\begin{split}
\overline{\ell'_{L}}&=\overline{\ell_{L}}V_L^\dagger, \\
(\ell^{\prime c})_L&=V_R^* (\ell^{c})_L, \\
\overline{(\ell^{\prime c})_L}&=\overline{(\ell^{c})_L} V_R^T.
\end{split}
\end{equation}
\footnote{From now on, to ease notation, it is left understood that the charge conjugation operation is to be performed before complex conjugation, \textit{i.e.}, $\bar{\ell_a^c} := \overline{\ell_a^c}$.} With this we have, for mass eigenstates
\begin{equation}
\mathcal{L}_{U\ell\ell}=\sum_{a,b} g_U\bar{\ell_a^c}\gamma^\mu P_L (V_U)_{ab} \ell_b\thinspace U_\mu^{++} + g_U\bar{\ell_a}\gamma^\mu P_L (V_U^\dagger)_{ab} \ell_b^c\thinspace U_\mu^{--},
\end{equation}
where $V_U \equiv V_R^T V_L$ is a unitary matrix. However, because both fermions in any term above carry the same conserved charge, there are two terms that contribute to any vertex. We may transpose one of them to, after carrying the spinor and charge conjugation algebra, rewrite the Lagrangian as
\begin{equation}
\begin{split}
\mathcal{L}_{U\ell\ell}&=\sum_{b> a} g_U \left\{\bar{\ell_a^c}\gamma^\mu [P_L (V_U)_{ab}-P_R (V_U)_{ba}] \ell_b\thinspace U_\mu^{++}
+\bar{\ell_a}\gamma^\mu [P_L (V_U^\dagger)_{ab}-P_R (V_U^\dagger)_{ba}]\ell_b^c \thinspace U_\mu^{--}\right\} \\
&+\sum_{a} g_U \left\{\bar{\ell_a^c}\gamma^\mu [P_L (V_U)_{aa}]\ell_a\thinspace U_\mu^{++}
+\bar{\ell_a}\gamma^\mu [P_L (V_U^\dagger)_{aa}] \ell_a^c\thinspace U_\mu^{--}\right\} ,
\end{split}
\end{equation}
with $a,b=e,\mu,\tau$. Notice that we have chosen to leave the heaviest lepton on the right. This exact manipulation and the resulting vertices are a common source of confusion.

Although $g_U$ is, in general, a free parameter, if \textit{(i)} the symmetry breaking pattern of the underlying model is such that $SU(2)_L \subset G$, where $G$ is a simple group, and \textit{(ii)} $U^{\pm \pm}$ is a maximally mixed combination of $N$ of its gauge bosons, then $g_U=\frac{g_{2L}}{\sqrt{N}}$, where $g_{2L}$ is the Standard Model $SU(2)_L$ gauge coupling. This follows from a matching condition at the breaking scale which dictates $g_G=g_{2L}$ \cite{Georgi:1977wk}. Although this construction may seem to be a strong imposition, the charged vector bosons of many common models tend to satisfy these requirements as the theories generalize the standard model electric charge scheme in such a way that the adjoint charge eigenstates are proportional to $\frac{T_i \pm i T_j}{\sqrt{2}}$, where $T_{i,j}$ are two generators of the gauge algebra. We use this to completely fix $g_U$ numerically, selecting the case where $U^{\pm \pm}$ is a combination of 2 gauge bosons, and write $g_U=\frac{g_{2L}}{\sqrt{2}}$. This construction corresponds, for instance, precisely to the case of the 3-3-1 model, but should cover a considerable sector of theory space. We write the final  $U\ell\ell$ effective Lagrangian
\begin{equation}\label{eq:Uint}
\begin{split}
\mathcal{L}_{U\ell\ell}&=\sum_{b> a} \frac{g_{2L}}{\sqrt{2}} \left\{\bar{\ell_a^c}\gamma^\mu [P_L (V_U)_{ab}-P_R (V_U)_{ba}] \ell_b\thinspace U_\mu^{++} + \bar{\ell_a}\gamma^\mu [P_L (V_U^\dagger)_{ab}-P_R (V_U^\dagger)_{ba}] \ell_b^c\thinspace U_\mu^{--}\right\} \\
&+ \sum_{a} \frac{g_{2L}}{\sqrt{2}} \left\{\bar{\ell_a^c}\gamma^\mu [P_L (V_U)_{aa}] \ell_a\thinspace U_\mu^{++} +\bar{\ell_a}\gamma^\mu [P_L (V_U^\dagger)_{aa}] \ell_a^c\thinspace U_\mu^{--}\right\} .
\end{split}
\end{equation}
From which the interaction between same flavor leptons may be perceived to be purely axial: $\bar{\ell_a^c}\gamma^\mu P_L  \ell_a=-\bar{\ell_a^c}\gamma^\mu \frac{\gamma^5}{2}  \ell_a$.

Now we turn to the doubly-charged scalar. While the previous interaction was originated by the higher symmetry covariant derivative, this one comes from Yukawa Lagrangians. We have as the most general effective interaction
\begin{equation}\label{eq:Yint1}
\mathcal{L}_{Y\ell\ell}=-\sum_{a,b}g_{YL}\left\{ \bar{\ell^c_a}(\mathcal{O}_Y)_{ab} P_L \ell_b\thinspace Y^{++} + \bar{\ell_a}(\mathcal{O}_Y^\dagger)_{ab} P_R \ell_b^c\thinspace Y^{--} \right\}.
\end{equation}
In the Lagrangian above, the fermions are already mass eigenstates and the interaction mixing matrix is arbitrary: it is related to one a priori (arbitrary) Yukawa matrix $G_Y$ as $\mathcal{O}_Y=V_R^T G_{Y} V_L$. Again, it is not necessary to add a second handedness term.

We can once more work with both terms in Eq.~(\ref{eq:Yint1}) that involve a pair $a,b$ of (equal charge) leptons in order to arrange them into an identical spinor chain, after which we obtain the useful form Lagrangian:
\begin{equation}\label{eq:Yint2}
\begin{split}
\mathcal{L}_{Y\ell\ell}=&-\sum_{b>a}g_{YL}\left\{ \bar{\ell^c_a}\left[(\mathcal{O}_Y)_{ab}+(\mathcal{O}_Y)_{ba}\right] P_L \ell_b\thinspace Y^{++} + \bar{\ell_a}\left[(\mathcal{O}_Y^\dagger)_{ab}+(\mathcal{O}_Y^\dagger)_{ba}\right] P_R \ell_b^c\thinspace Y^{--} \right\} \\
&-\sum_{a=b}g_{YL}\left\{ \bar{\ell^c_a}\left[(\mathcal{O}_Y)_{aa}\right] P_L \ell_a\thinspace Y^{++} + \bar{\ell_a}\left[(\mathcal{O}_Y^\dagger)_{aa}\right] P_R \ell_a^c\thinspace Y^{--} \right\}.
\end{split}
\end{equation}

Lastly, we write the neutral scalar interaction Lagrangian. Lorentz and electric charge invariance dictates it must be simply
\begin{equation}
\begin{split}
\mathcal{L}_{s\ell\ell}&= - g_{sL}\bar{\ell}\mathcal{O}_s P_L \ell\thinspace s
-g_{sL}\bar{\ell}\mathcal{O}_s^\dagger P_R \ell\thinspace s\\
&=-\sum_{a,b}g_{sL}\bar{\ell_a}\left[(\mathcal{O}_s)_{ab} P_L+(\mathcal{O}^\dagger_s)_{ab} P_R\right] \ell_b\thinspace s,
\end{split}
\end{equation}
where $\mathcal{O}_s$ is arbitrary and related to a Yukawa matrix as $\mathcal{O}_s=V_R^\dagger G_{s} V_L$.

Obviously, the appropriate effective model also contains kinetic terms defined by
\begin{equation}
\begin{split}
\mathcal{L}_{\text{kin}}=&-\frac{1}{2}U_{\mu\nu}^\dagger U^{\mu\nu}+M_U^2 (U^{++})^\dagger U^{++} \\
&+(\partial_\mu Y^{++})^\dagger \partial^\mu Y^{++} - M_Y^2 (Y^{++})^\dagger Y^{++} \\
&+ \frac{1}{2}\partial_\mu s\thinspace \partial^\mu s - \frac{1}{2}M_s^2\thinspace s^2,
\end{split}
\end{equation}
where $U_{\mu\nu}=\partial_\mu U^{++}_\nu-\partial_\nu U^{++}_\mu$.

Three 2-particle scenarios will be considered, each with a pair of exotic species which interfere. The correspondent Lagrangians are

\begin{equation}\label{eq:lagint}
\begin{split}
\mathcal{L}_{U-s}&=\mathcal{L}_{\text{kin}}+\mathcal{L}_{U\ell\ell}+\mathcal{L}_{s\ell\ell}, \\
\mathcal{L}_{U-Y}&=\mathcal{L}_{\text{kin}}+\mathcal{L}_{U\ell\ell}+\mathcal{L}_{Y\ell\ell}, \\
\mathcal{L}_{Y-s}&=\mathcal{L}_{\text{kin}}+\mathcal{L}_{Y\ell\ell}+\mathcal{L}_{s\ell\ell}.
\end{split}
\end{equation}

With a general, model-independent parametrization of the needed interactions, we may return to the central objective regarding which a few comments are in order. It is true that if the matrix elements of the 3 mixing matrices parametrizing the interaction Lagrangians could be arbitrarily small, any experimental constraint could be easily met; however, if these particles do exist \textit{(i)} elements too small are not desirable because of matters such as naturalness and; \textit{(ii)} more importantly, the theoretically predicted unitarity of the $V_U$ matrix is powerful with respect to inducing exclusion contours.

Concerning the parameter space, notice that $g_s$ and $g_Y$ could be absorbed into their corresponding mixing matrices, and although we write them explicitly on analytical expressions (mostly for book keeping purposes), they will be effectively set to $1$ in all numerical evaluations. Notice also, checking Eq. (\ref{eq:Yint2}), that any element of $\mathcal{O}_Y$ only appears together with its symmetric partner, so that this effective mixing matrix may be taken symmetric. The free parameters in each scenario, which include masses and degrees of freedom of the applicable matrix, may then be checked to be as appears on Table~\ref{Tab:nmbdof}.

Since to effect numerical optimization with the number of free parameters that exists when considering the general case is impractical, we considerably reduce the number of parameters by restricting the analysis to real matrices. The $V_U$ unitary matrix then becomes an orthogonal one, whose determinant may be chosen to be $1$ without loss of generality, and which we parametrize with Euler Angles

\begin{table}[t!]
\caption{Number of parametric degrees of freedom contained within each scenario if the mixing matrices are regarded as complex and real.}\label{Tab:nmbdof}
\begin{tabular}{ccc}
\toprule
Scenario    &    Complex    &    Real         \\ \midrule
$U-Y$       &       23      &     11          \\
$U-s$       &       29      &     14          \\
$s-Y$       &       32      &     17          \\ \bottomrule
\end{tabular}
\end{table}

\begin{equation}
V_U = \begin{pmatrix}
\cos\psi\cos\phi-\cos\theta\sin\phi\sin\psi & \cos\psi\sin\phi+\cos\theta\cos\phi\sin\psi & \sin\theta\sin\psi \\
-\sin\psi\cos\phi-\cos\theta\sin\phi\cos\psi & -\sin\psi\sin\phi+\cos\theta\cos\phi\cos\psi & \sin\theta\cos\psi \\
\sin\theta\sin\phi & -\sin\theta\cos\phi & \cos\theta
\end{pmatrix}.
\end{equation}

Lastly, we should clarify the role of this matrix, considering the complete, unitary case. Referring to $V_U \equiv V_R^T V_L$, there are two situations in which this mixing can be ignored in a natural way: \textit{(i)} If $V_R$ and $V_L$ could be set to $\mathbb{1}$. This can occur whenever the mass matrix and every leptonic interaction can be simultaneously diagonalized, which is not the general case and relates to a small part of theory space. \textit{(ii)} An alternative independent possibility is $V_R = {V_L}^*$. This implies that the mass matrix is diagonalized by an orthogonal transformation instead of by a biunitary one (notice, in particular, the condition above can only be met by real rotation matrices). A symmetric squared mass matrix of this type is, again, a special case. We consider a non-diagonal orthogonal $V_U$, which, apart from the missing phases, should be consistent with the general case.

\section{Calculations}
\label{sec:compmet}

\begin{figure}[t!]
	{\centering
	\includegraphics[width=0.4\linewidth]{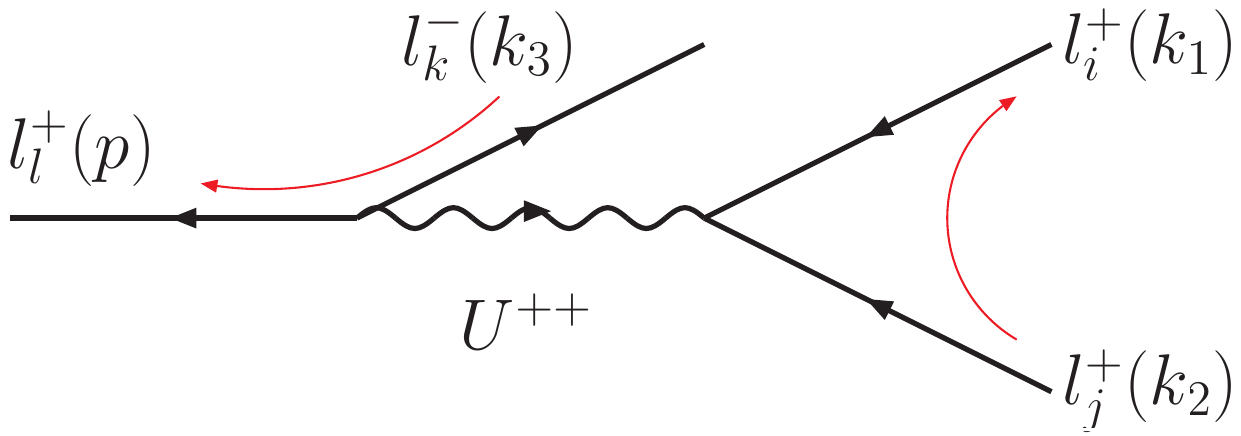}
	\hspace*{5mm}
	\includegraphics[width=0.4\linewidth]{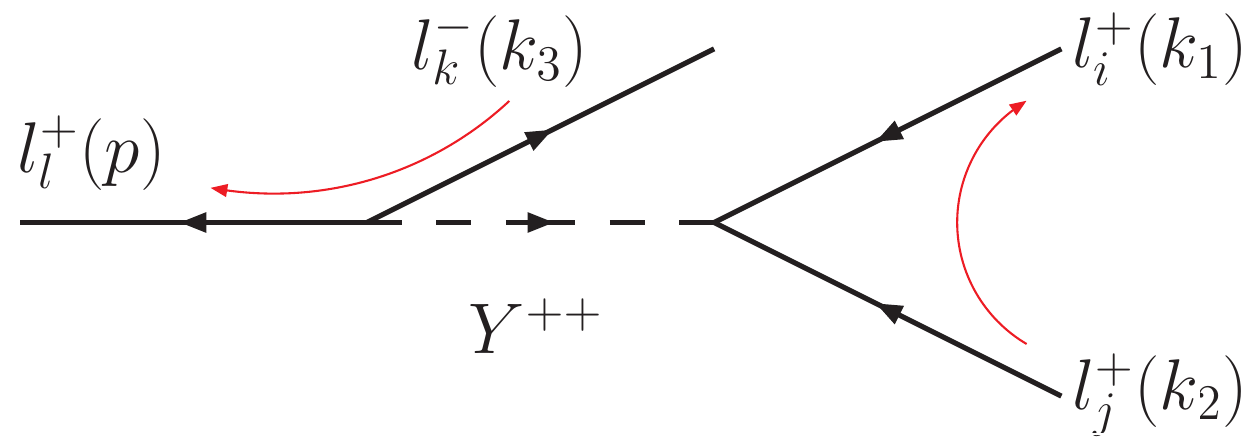} \\
	\vspace*{5mm}
	\includegraphics[width=0.4\linewidth]{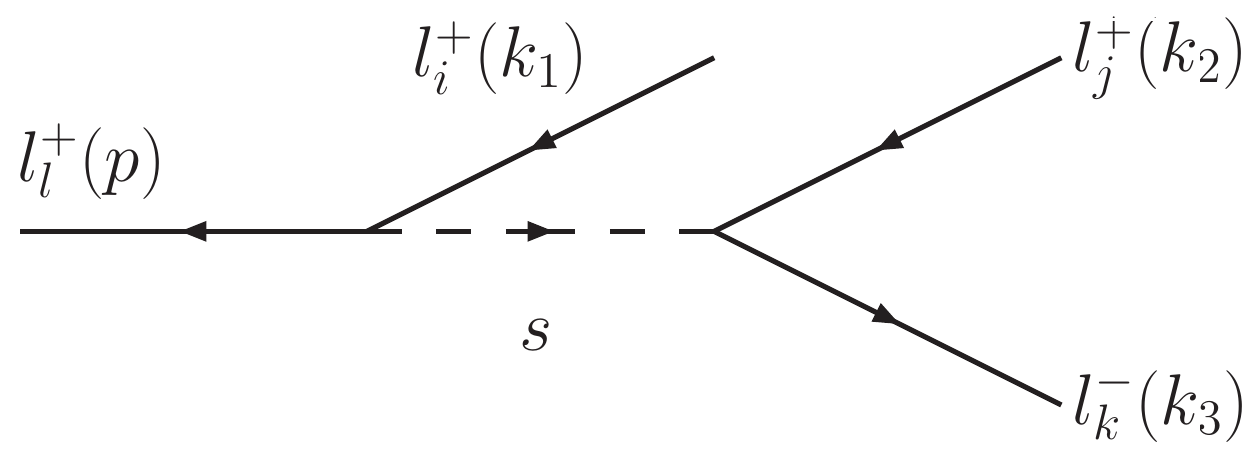}
	\hspace*{5mm}
	\includegraphics[width=0.4\linewidth]{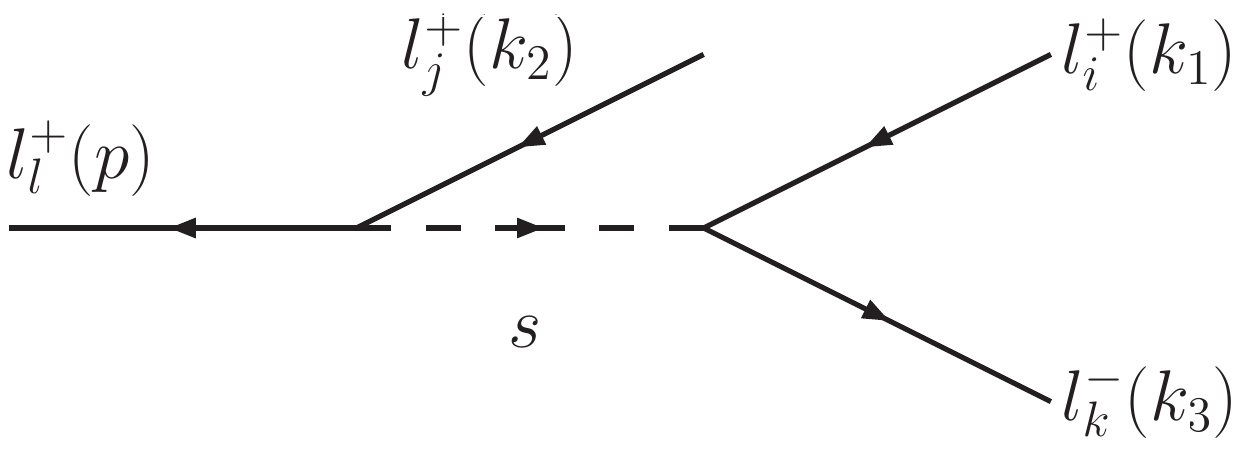}
	\caption{Explicit diagrams contributing to the 3-body lepton decays.}
	\label{fig:diagrams}
	}
\end{figure}

The diagrams contributing to a generic $\ell^+_l(p)\to \ell^+_i(k_1)\ell^+_j(k_2)\ell^-_k(k_3)$ decay appear on Figure~\ref{fig:diagrams}. The corresponding amplitudes are given by

\begin{equation}\label{eq:firstamp}
\begin{split}
i\mathcal{M}_U=\left(\frac{i g_{2L}}{\sqrt{2}}\right)^2&\bar{v}_{\ell_l}(p)\gamma^\mu (V_{Ukl}P_L-V_{Ulk}P_R)v_{\ell_k}(k_3)\frac{-i g_{\mu \nu}}{(k_1+k_2)^2-M_U^2} \\
\times&\bar{u}_{\ell_i}(k_1)\gamma^\nu (V_{Uij}P_L-V_{Uji}P_R)v_{\ell_j}(k_2),
\end{split}
\end{equation}

\begin{equation}
\begin{split}
i\mathcal{M}_Y=\left(-i g_Y\right)^2&\bar{v}_{\ell_l}(p)(\mathcal{O}_{Ylk}+\mathcal{O}_{Ykl})P_R v_{\ell_k}(k_3)\frac{i}{(k_1+k_2)^2-M_Y^2} \\
\times&\bar{u}_{\ell_i}(k_1) (\mathcal{O}_{Yij}+\mathcal{O}_{Yji})P_L v_{\ell_j}(k_2), \\
\end{split}
\end{equation}

\begin{equation}
\begin{split}
i\mathcal{M}_{s1}=(-1)\left(-i g_s\right)^2&\bar{v}_{\ell_l}(p)(\mathcal{O}_{sli}P_L+\mathcal{O}_{sil}P_R) v_{\ell_i}(k_1)\frac{i}{(k_2+k_3)^2-M_s^2} \\
\times&\bar{u}_{\ell_k}(k_3) (\mathcal{O}_{skj}P_L+\mathcal{O}_{sjk}P_R) v_{\ell_j}(k_2), \\
\end{split}
\end{equation}

\begin{equation}\label{eq:lastamp}
\begin{split}
i\mathcal{M}_{s2}=\left(-i g_s\right)^2&\bar{v}_{\ell_l}(p)(\mathcal{O}_{slj}P_L+\mathcal{O}_{sjl}P_R) v_{\ell_j}(k_2)\frac{i}{(k_1+k_3)^2-M_s^2} \\
\times&\bar{u}_{\ell_k}(k_3) (\mathcal{O}_{ski}P_L+\mathcal{O}_{sik}P_R) v_{\ell_i}(k_1).\\
\end{split}
\end{equation}
A didactic discussion on how to achieve this expressions from the Lagrangians in Eq.~(\ref{eq:lagint}) is presented on Appendix~\ref{app:amps}. Nevertheless, we cross check the amplitudes above by generating them through \texttt{FeynRules} \cite{Alloul:2013bka} in association with the \texttt{FeynArts} \cite{Hahn:2000kx} package.

\begin{table}[]
\caption{Current experimental limits on every channel of 3-body lepton decay.}\label{Tab:explim}
\begin{tabular}{lc}
\toprule
\multicolumn{1}{c}{Process}     &  BR \\ \midrule
$\mu^+  \to  e^+ e^- e^+$       &  $<1.0\times 10^{-12}$  \\
$\tau^+ \to  e^+ e^- e^+$       &  $<2.7\times 10^{-8}$   \\
$\tau^+ \to  e^+ \mu^- \mu^+$   &  $<2.7\times 10^{-8}$   \\
$\tau^+ \to \mu^+ e^- e^+$      &  $<1.8\times 10^{-8}$   \\
$\tau^+ \to \mu^+ \mu^- \mu^+$  &  $<2.1\times 10^{-8}$   \\
$\tau^+ \to \mu^+ e^- \mu^+$    &  $<1.7\times 10^{-8}$   \\
$\tau^+ \to  e^+ \mu^- e^+$     &  $<1.5\times 10^{-8}$   \\ \bottomrule
\end{tabular}
\end{table}

The current experimental limits on 3-body lepton decays are shown on Table~\ref{Tab:explim} (see \cite{ParticleDataGroup:2020ssz} and references therein). What we call \textit{solutions} are any sets of numbers identified with the free parameters which cause the branching ratios to obey the constraints.

The range for the non-mass parameters are as follows

\begin{equation}
\begin{split}
0& \leq \phi,\ \psi < 2\pi, \\
0& \leq \theta < \pi, \\
-1& < \mathcal{O}_{Yij},\ \mathcal{O}_{sij} < 1,
\end{split}
\end{equation}
which are chosen to exhaust the $V_U$ space and keep the scalar interactions perturbative. Moreover, because masses smaller than it are extremely unlikely and because the verification of this possibility wouldn't change the qualitative results of our analysis, we limit ourselves to masses above $\SI{500}{GeV}$.

The solutions are obtained through a simple constrained global optimization routine, repeated for more than 100 random seeds to verify the stability of the best results.

\begin{figure}[t!]
\adjustbox{center=10cm}{
	\makebox[1.2\linewidth]{
	\begin{subfigure}{0.5\linewidth}
	\centering
		\includegraphics[width=\linewidth,trim=0cm 0.4cm 0.8cm 1.2cm,clip]{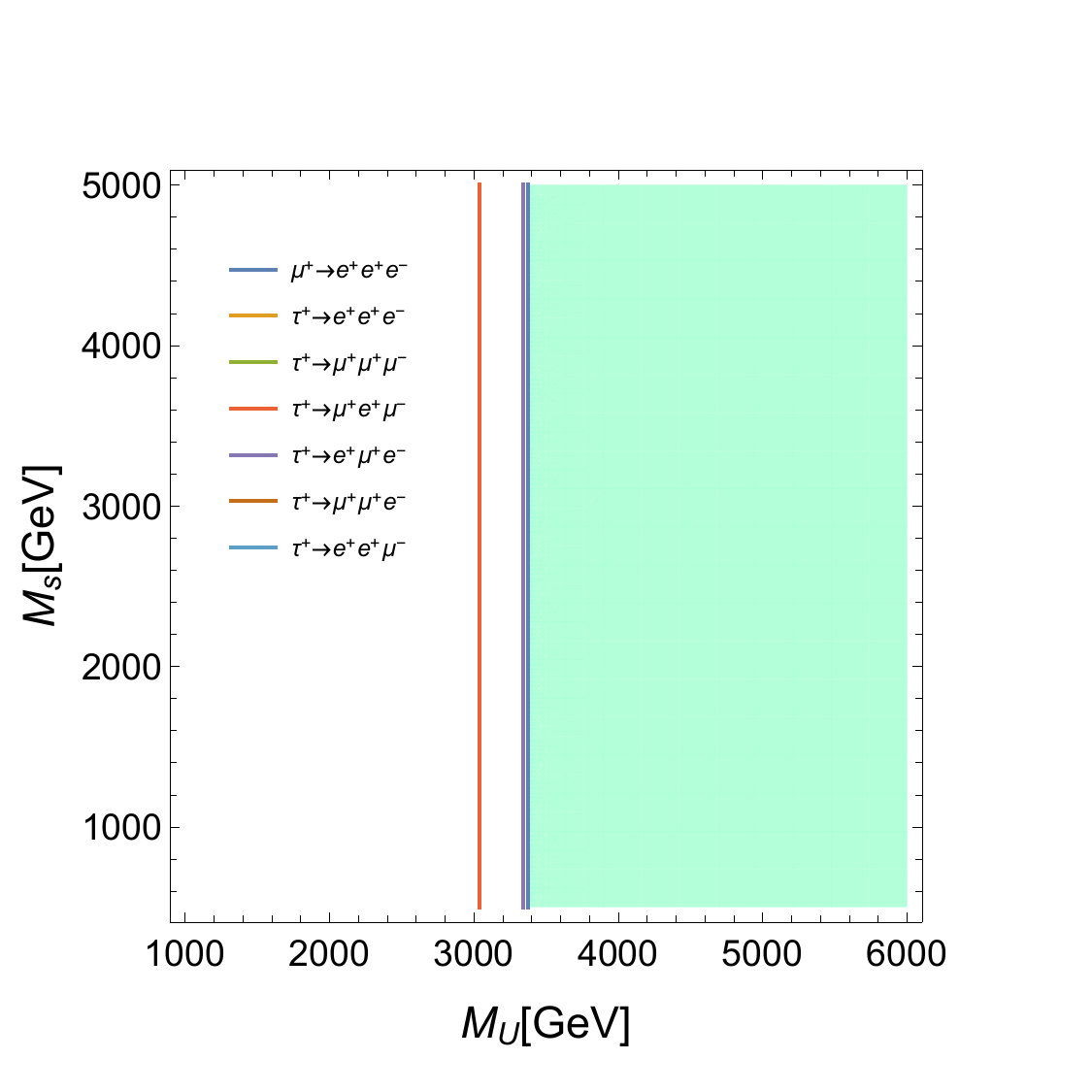}
		\caption{$|V_{Uij}|>10^{-3}$}\label{fig:contoursUa}
	\end{subfigure}
	\begin{subfigure}{0.5\linewidth}
	\centering
		\includegraphics[width=\linewidth,trim=0cm 0.4cm 0.8cm 1.2cm,clip]{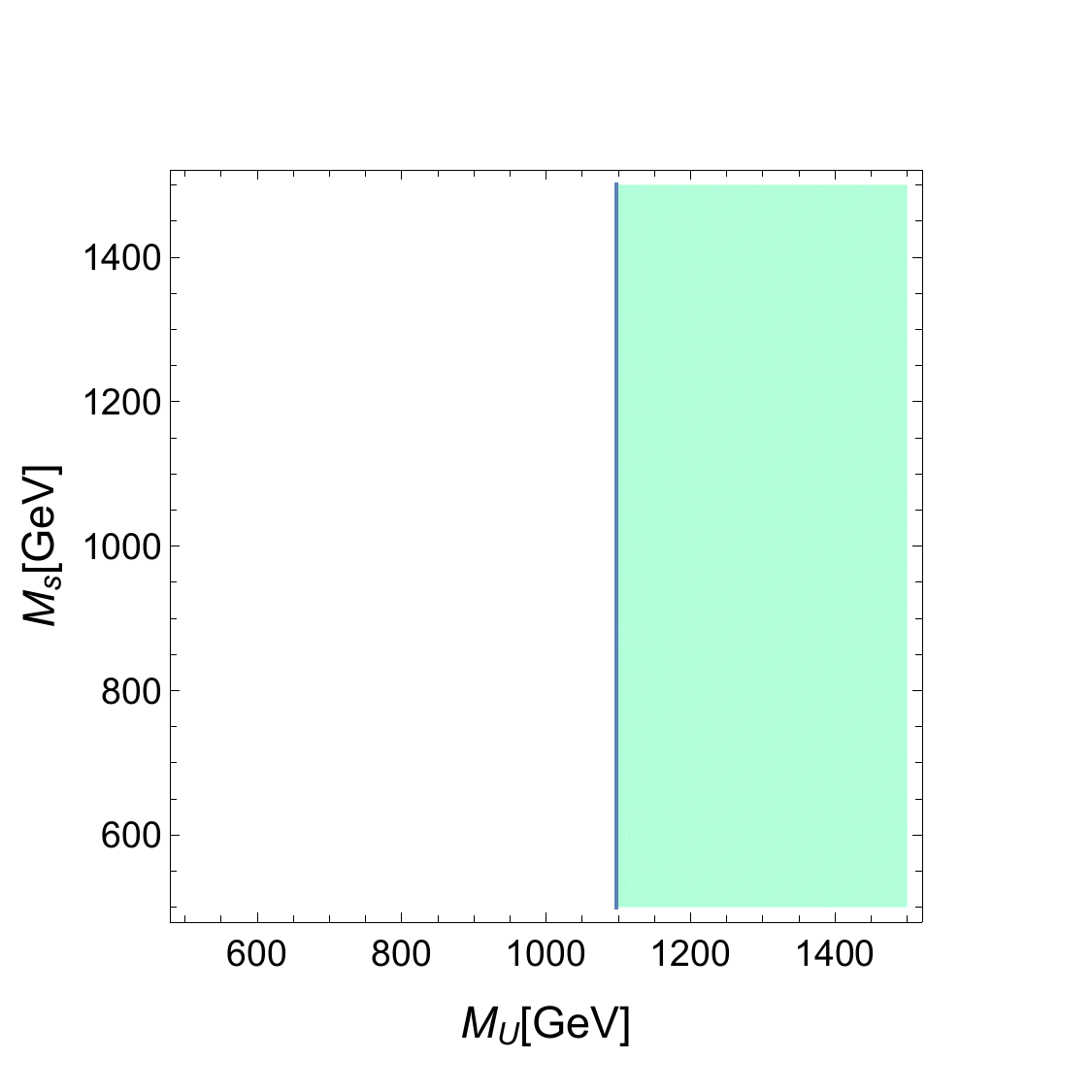}
		\caption{$|V_{Uij}|>10^{-4}$}\label{fig:contoursUb}
	\end{subfigure}}}  \\
	\vspace*{5mm}
\adjustbox{center=10cm}{
	\makebox[1.2\linewidth]{
	\begin{subfigure}{0.5\linewidth}
	\centering
		\includegraphics[width=\linewidth,trim=0cm 0.4cm 0.8cm 1.2cm,clip]{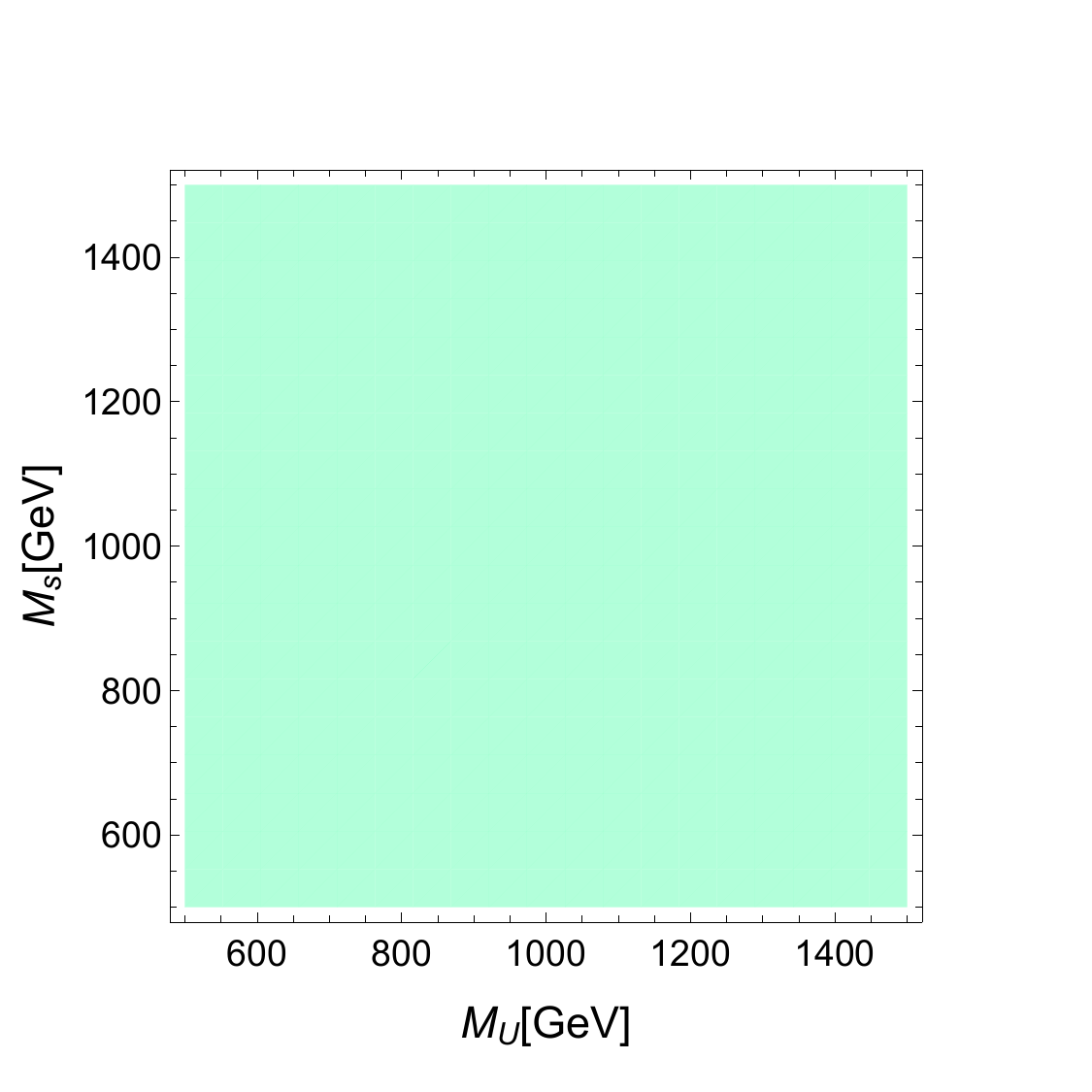}
		\caption{$|V_{Uij}|>10^{-5}$}
	\end{subfigure}
	\begin{subfigure}{0.5\linewidth}
	\centering
		\includegraphics[width=\linewidth,trim=0cm 0.4cm 0.8cm 1.2cm,,clip]{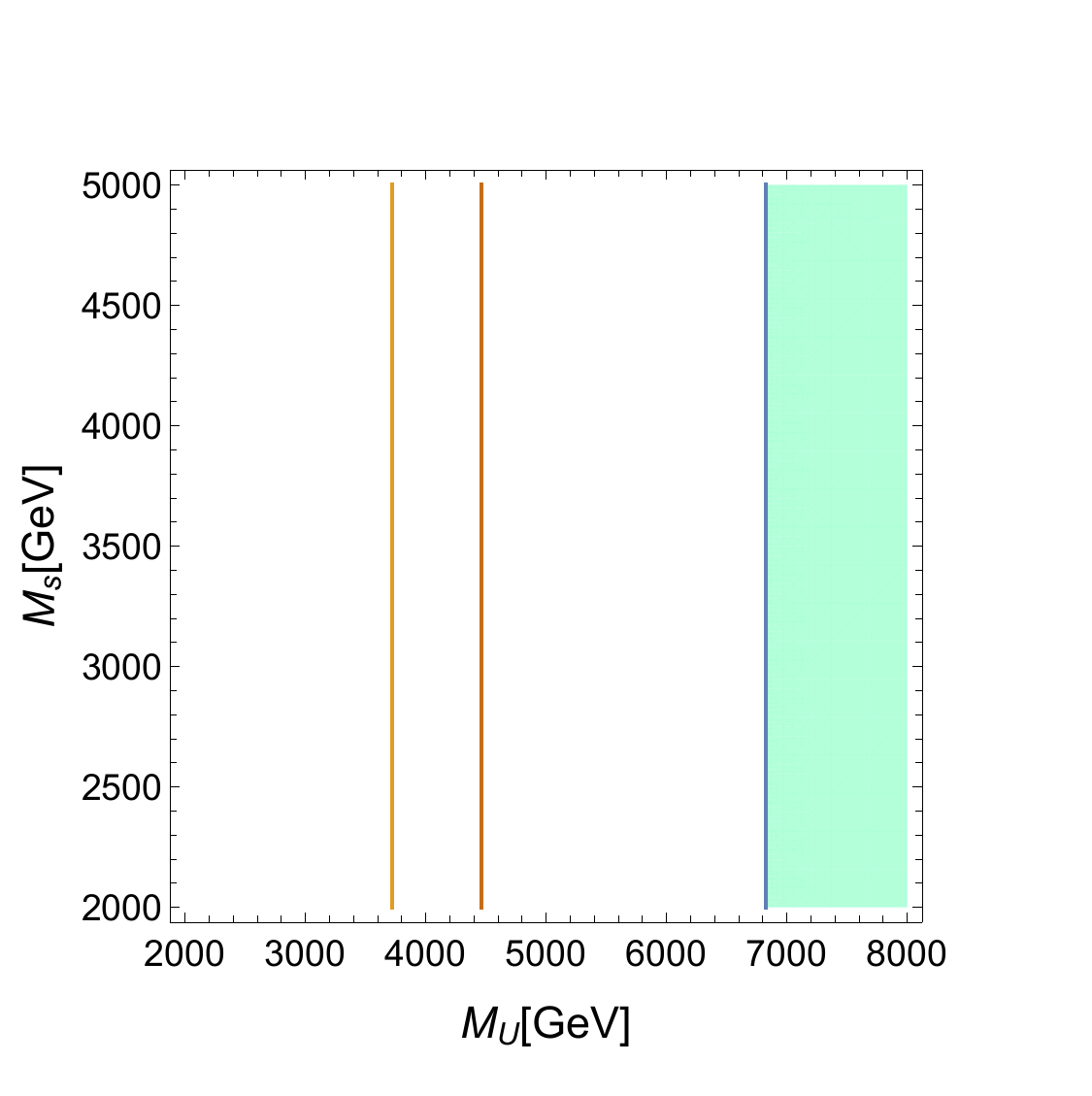}
		\caption{$|V_{Uij}|>10^{-5}$ and $V_{11}=0.85$}\label{fig:contoursUd}
	\end{subfigure}}}
	\caption{Exclusion ranges for $M_U$ generated by the experimental bounds on the various branching rations for leptonic decays, for each of our 4 benchmark cases.  The allowed region is painted green.}
	\label{fig:contoursU}
\end{figure}

\section{Results and analysis}
\label{sec:res}

In the scenarios where the $U^{\pm \pm}$ is present, we search for solutions which prioritize its mass -- i.e., we seek sets of numbers which minimize $M_U$, with every other parameter, including $M_s,M_Y$, free. We make this choice because vector bosons usually and for a greater part of parameter space impose stronger constraints, but, specially, because the predicted unitarity of the bilepton mixing sets it apart phenomenologically and makes it a hardly constrained field. The achievement and examination of these solutions is the operational objective of this work.

Besides what has already been discussed, we consider additional benchmark conditions to fix the lower bound on the modulus of matrix elements -- because of reasons exposed in the previous section. These conditions are detailed in the Figures and in the subsections below.

In addition, we consider benchmark impositions on the diagonal couplings of the mixing matrices. Also very restrictive, these constraints are designed to check what the lowest possible masses are if the model that correctly describes nature couples equal flavor particles (almost) maximally.

\begin{figure}[t!]
\adjustbox{center=10cm}{
	\makebox[1.2\linewidth]{
	\begin{subfigure}{0.5\linewidth}
	\centering
		\includegraphics[width=\linewidth,trim=0cm 0.4cm 0.8cm 1.2cm,clip]{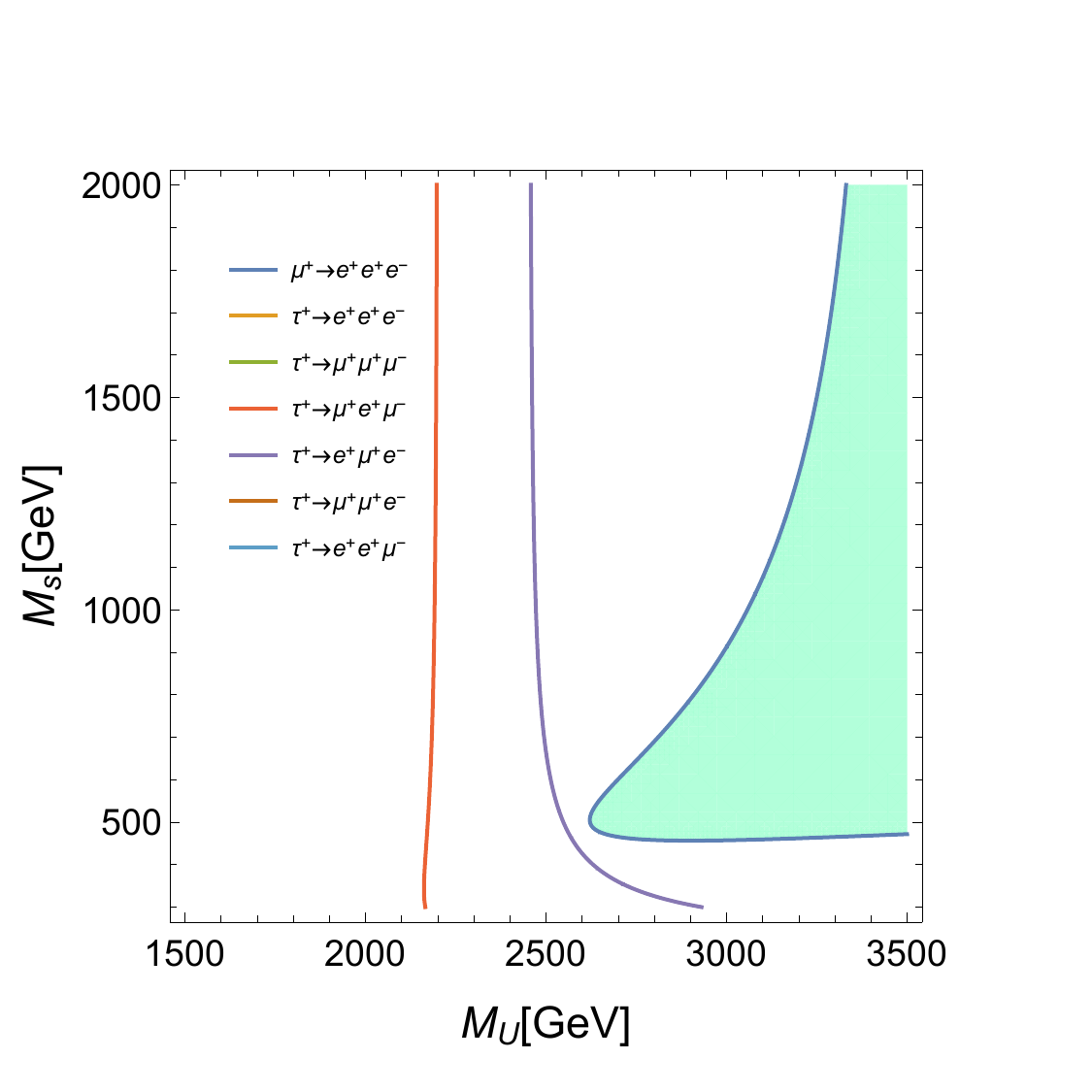}
		\caption{$|O_{sij},V_{Uij}|>10^{-3}$}
	\end{subfigure}
	\begin{subfigure}{0.5\linewidth}
	\centering
		\includegraphics[width=\linewidth,trim=0cm 0.4cm 0.8cm 1.2cm,,clip]{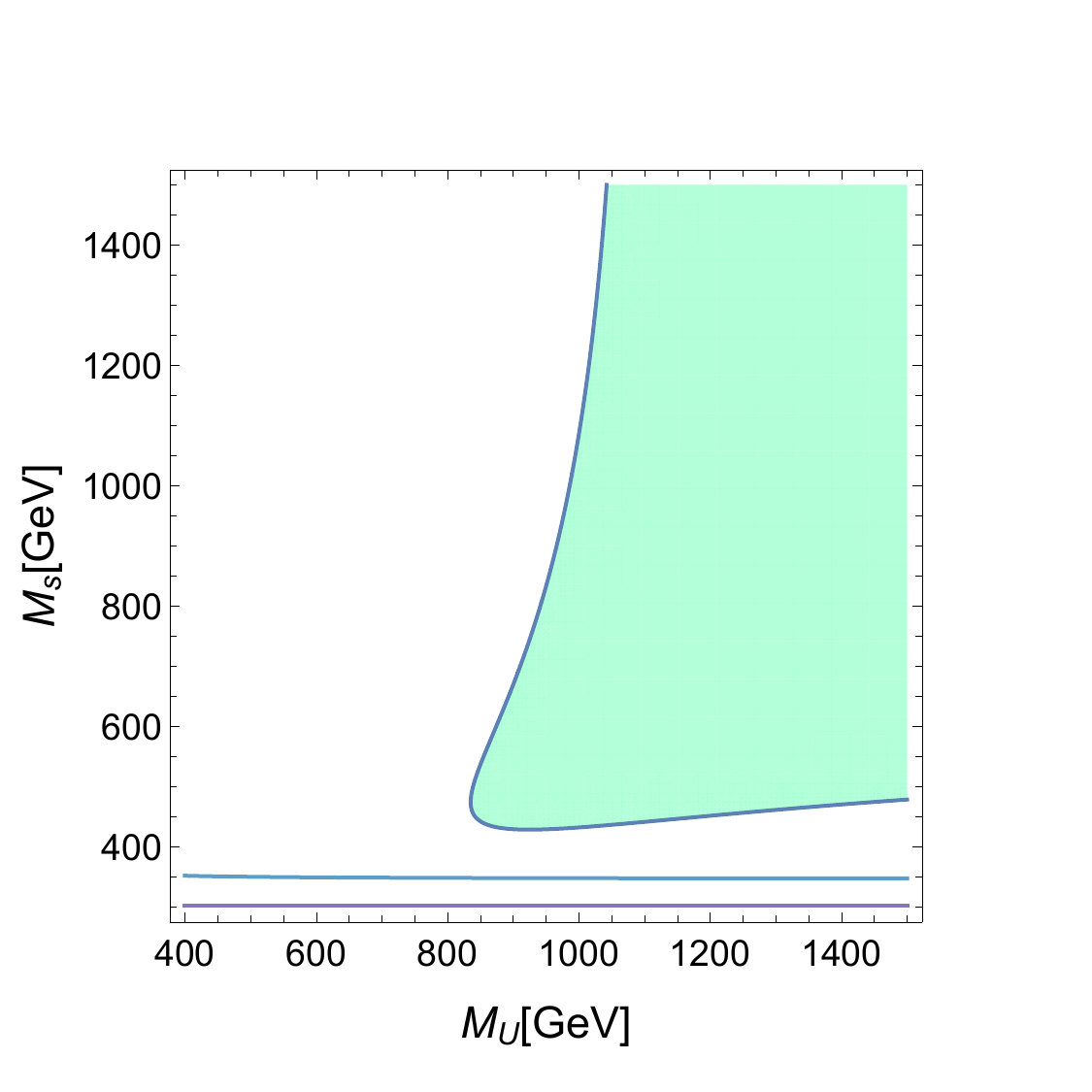}
		\caption{$|O_{sij},V_{Uij}|>10^{-4}$}
	\end{subfigure}}}  \\
	\vspace*{5mm}
\adjustbox{center=10cm}{
	\makebox[1.2\linewidth]{
	\begin{subfigure}{0.5\linewidth}
	\centering
		\includegraphics[width=\linewidth,trim=0cm 0.4cm 0.8cm 1.2cm,,clip]{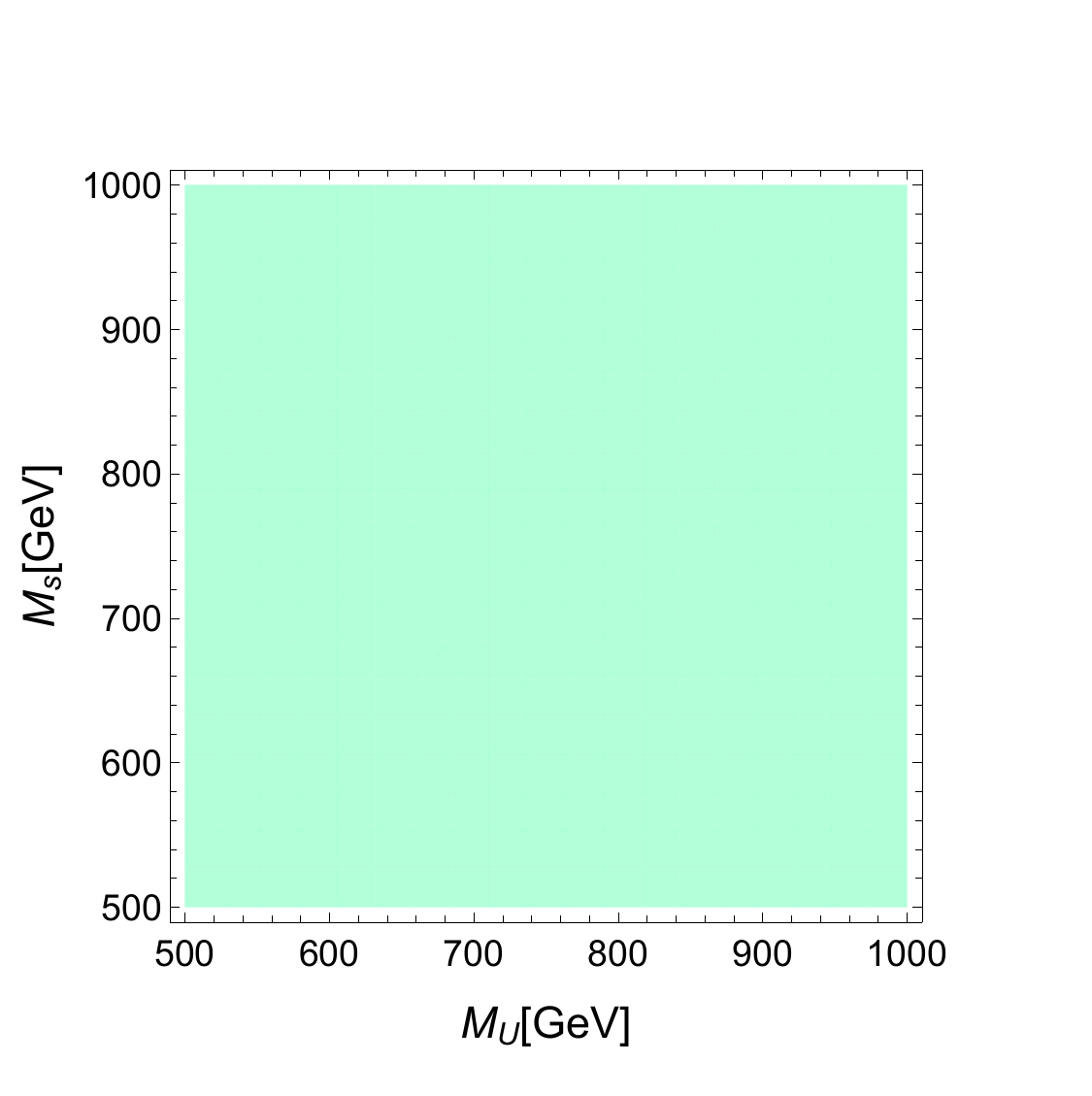}
		\caption{$|O_{sij},V_{Uij}|>10^{-5}$}
	\end{subfigure}
	\begin{subfigure}{0.5\linewidth}
	\centering
		\includegraphics[width=\linewidth,trim=0cm 0.4cm 0.8cm 1.2cm,,clip]{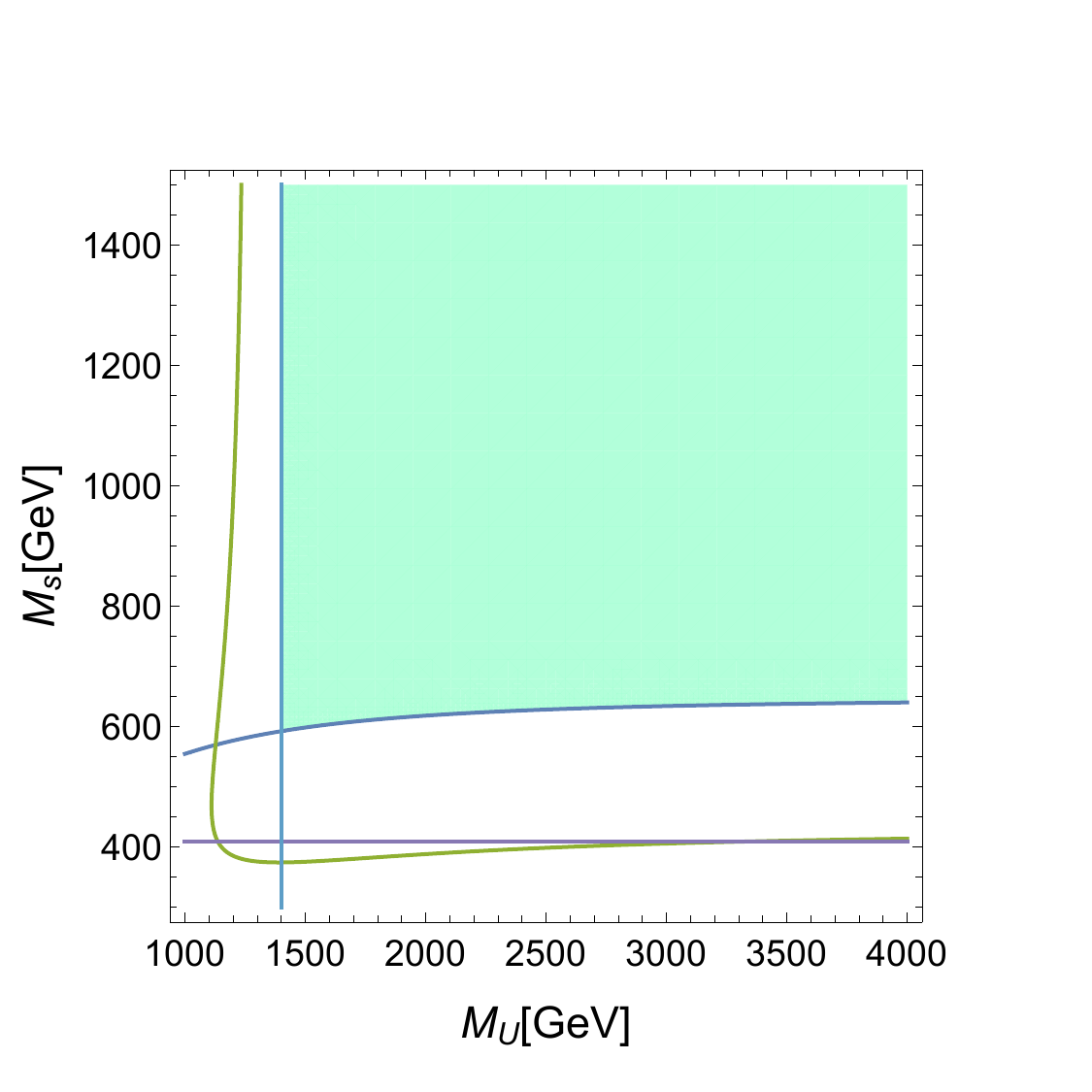}
		\caption{$|O_{sij},V_{Uij}|>10^{-5}$ and $O_{s11}=O_{s22}=1$
		}	\label{fig:contoursUsd}
	\end{subfigure}}}
	\caption{Exclusion contours on the $M_U \times M_s$ plane generated by the bounds on 3-body Lepton Flavor Violating decays. The allowed region is painted green.}
	\label{fig:contoursUs}
\end{figure}

\subsection{Pure $U$ Scenario}
\label{Sec:U}

We begin examining the constraints resulting from a doubly-charged vector bilepton alone. The results are presented in Figure~\ref{fig:contoursU} and the solutions in Table~\ref{Tab:solU}. We show the contours in the $M_U \times M_s$ plane even though there is no $M_s$ dependence to facilitate comparison with the scenario below.

We recognize that to allow for masses of the order of $\SI{1100}{GeV}$ (see Fig.~\ref{fig:contoursUb}) we need an hierarchy\footnote{Since, for an orthogonal matrix, small elements imply the need for large ones.} within $V_U$ already similar to that of the Yukawa sector for quarks in the SM, such that we approach a non-natural parameter regime and is why five orders of magnitude is the largest hierarchy we allow. But the greatest feature to observe is that with small general hierarchy (Fig.~\ref{fig:contoursUa}) or with large \textit{but not maximal} diagonal coupling (Fig.~\ref{fig:contoursUd}) the constraints are strong, demanding $M_U> \SI{3200}{GeV}$ and $M_U> \SI{6900}{GeV}$, respectively.  It must be recognized that, in each instance, the contours result from a specific, not always evident, interplay between one or various BR bounds and the unitarity conditions of $V_U$.

To illustrate the complementarity between the LHC phenomenology and the CLFV 3-body decays analysis, consider Refs.~\cite{Dutta:1994pd,Dion:1998pw,Barreto:2013paa,RamirezBarreto:2011av,RamirezBarreto:2008wq,Nepomuceno:2016jyr,Tully:1999yg,Meirose:2011cs,Barela:2019pmo,Corcella:2018eib,Corcella:2017dns,Frampton:2020pef,Coriano:2018khp}. These studies focus on the 3-3-1 model, the only simple well known SM extension to contain a doubly-charged vector bilepton. In this special case, $V_U$ cannot be the identity and should not be ignored: the reason is that, even in the minimal version of the model, lepton masses arise from two distinct Yukawa sources, which precludes the possibility of their squared mass matrix being set diagonal from the start and, furthermore, the matrix is not symmetric. In the referred studies, $V_U$ is not treated, and flavor conserving processes are investigated with the diagonal matrix elements implicitly fixed to 1. Since this cannot be exact within the 3-3-1, this choice corresponds to the case of high hierarchy $|V_{Uij}|>10^{-5}$ -- to validate this claim it should also be checked that our corresponding solution is of the form $V_U \sim \mathbb{1}$, which is indeed the case. 

The strongest bounds from the mentioned literature are capable of excluding bilepton masses $M_U \lesssim \SI{1}{TeV}$. The implication is that for the sector of theory space with a $V_U$ hierarchy of $10^{4}$ or lower, the CLFV lepton decays bounds should be considered, while for the flavor conserving sector, numerically and casually equivalent to tolerant hierarchies of $10^{5}$ or higher, the more energetic LHC phenomenology should be more appropriate.

Notice that the argument above is general: for not exceedingly low, improbable masses, the case of no mixing, $V_U = \mathbb{1}$ (which can be contained in a natural way within a theory as discussed in Sec.~\ref{sec:effint}), in which our processes do not occur at all, is well described by the most permissive case of high hierarchy $|V_{Uij}|>10^{-5}$. With the discussion carried in this section entirely in mind, we move on to the alternative scenarios.

\subsection{$U-s$ Scenario}

The scenario which includes a doubly-charged vector bilepton and a neutral scalar is the most complex and illuminating one, because it contains the unitary $V_U$ matrix and possibly strong interference. The exclusion contours are shown in Figure~\ref{fig:contoursUs} and the solutions appear at Table~\ref{Tab:solUs}.

\begin{figure}[t!]
	\centering
		\includegraphics[width=0.6\linewidth]{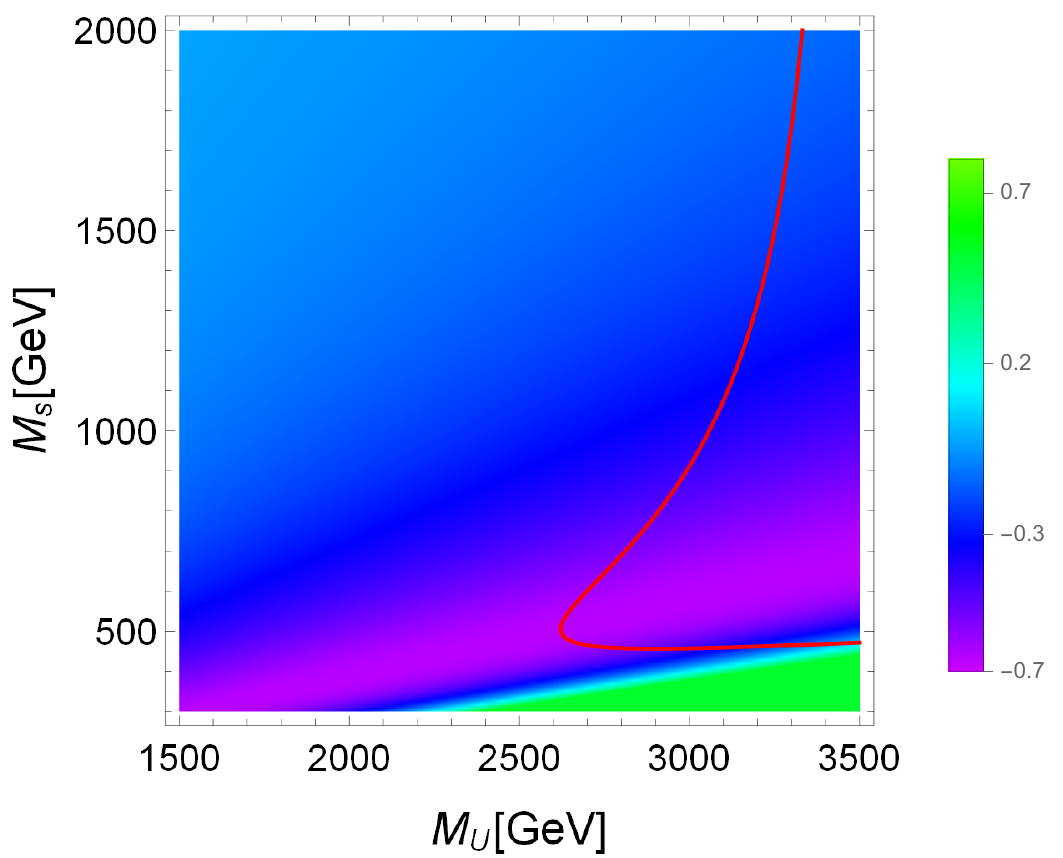}
		\caption{Density plot of the extra contributions to $\mu^+ \to e^+ e^- e^+$, caused by the addition of the neutral scalar $s$ to a model with the doubly-charged vector bilepton $U^{\pm \pm}$. }
	\label{fig:interference}
\end{figure}

From the contours, we learn that to allow for bilepton masses of the order of $M_U<\SI{1}{TeV}$, at least some effective couplings $g_{\text{eff}}\sim g_U V_{Uij}$ must be set as low as $<10^{-4}$, meanwhile the entire parameter space is possible if the matrix elements are allowed to become as small as $~10^{-5}$, showing, again, the complementarity between the phenomenology of CLFV lepton decays and LHC processes, which cover, respectively, the non-diagonal and diagonal $V_U$ models. 
Additionally, regarding the neutral scalar boson $s$, we observe, in Figure~\ref{fig:contoursUsd}, that if $\mathcal{O}_{s11}=\mathcal{O}_{s22}=1$ is enforced, the bound on $M_U$ is strengthened from\footnote{We stress that $\SI{500}{GeV}$ is simply the lowest mass point we consider.} $M_U>\SI{500}{GeV}$ to $M_U>\SI{1500}{GeV}$ while virtually unchanging the bound on $M_s$. This just reasserts that the vector contribution is indeed dominant, which, again, could be expected from the unitarity of $V_U$ and the fact that there is less possibilities in spin space for a scalar mediated process.

We also show, in Figure~\ref{fig:interference}, a density plot of the ratio
\begin{equation}
\frac{\text{BR}_{U-s}+\text{BR}_{s}}{\text{BR}_{U}},
\end{equation}
where $\text{BR}_{X-Y}$ is the contribution of the interference between $X$ and $Y$ and $\text{BR}_{X}$ is the pure $X$ contribution to the BR. This information is useful to investigate how the presence of a second particle may relieve naive constraints derived from single exotic particle Lagrangians.  We notice that, even if the scalar contribution is significantly smaller, it allows the solution to enhance destructive interference, which causes the distortion on the inferior left corner of the contour and contributes to, in the $|O_{sij},V_{Uij}|>10^{-3}$ case, rendering constraints softer by 19\% on $M_U$.

\subsection{$U-Y$ Scenario}

This construction is simpler and the results can be infered with reasoning alone.

The interference terms of the $U^{\pm \pm}$ with the $Y^{\pm \pm}$-particle in the $\mu^+ \to e^+ e^- e^+$ branching ratio are proportional to the electron mass, therefore interference effects are negligible to our purposes because $m_e$ is much smaller than the next mass scale. This indicates that possible solutions for this scenario involve parameters, related to the vector bilepton, identical to those of the pure $U^{\pm \pm}$ scenario, of Section \ref{Sec:U}, with $Y^{\pm \pm}$-related parameters very small in modulus, the least allowed by the benchmark conditions. This guarantees that the  $Y^{\pm \pm}$ contribution is rendered negligible and doesn't affect the exclusion contour, turned similar to those of Figure~\ref{fig:contoursU}.

\begin{figure}[t!]
	\centering
		\includegraphics[width=0.6\linewidth]{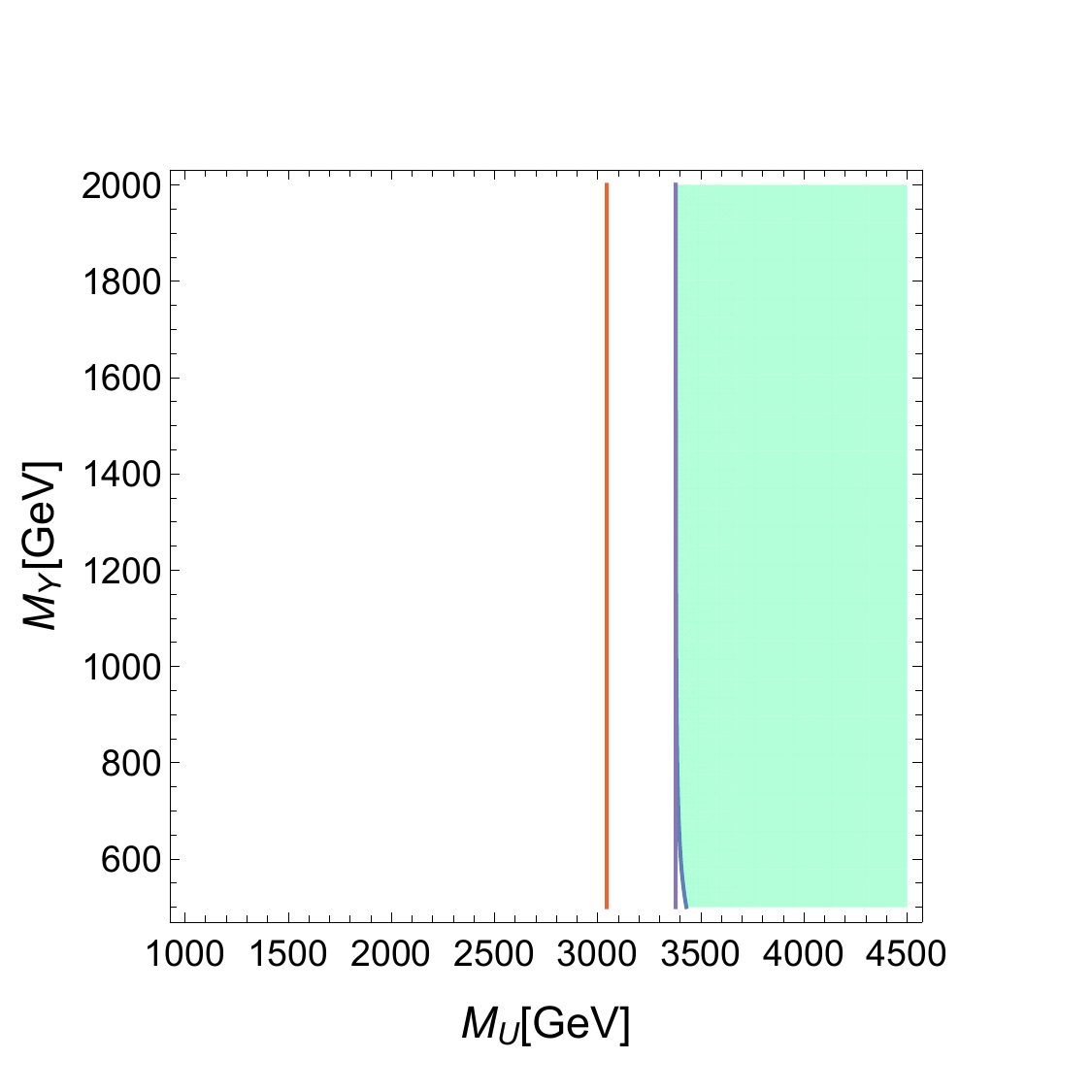}
		\caption{Exclusion contour on the $M_U \times M_Y$ plane, showing that the solution subjected to the $|O_{Yij},V_{Uij}|>10^{-3}$ condition which gives the weakest bounds on the $M_U$ mass is analogous to the one of the pure $U^{\pm \pm}$ case, with negligible scalar contributions (except when the mass of the scalar is exceedingly low, of course). }
	\label{fig:contoursUY}
\end{figure}

To illustrate the point above, we show, in Figure~\ref{fig:contoursUY} the plot corresponding to the solution of Fig.~\ref{fig:contoursUa} together with $O_{Yij}=10^{-3}$.

\subsection{$Y-s$ Scenario}

\begin{figure}[t!]
\adjustbox{center=10cm}{
	\makebox[1.2\linewidth]{
	\begin{subfigure}{0.5\linewidth}
	\centering
		\includegraphics[width=\linewidth,trim=0cm 0.4cm 0.8cm 1.2cm,clip]{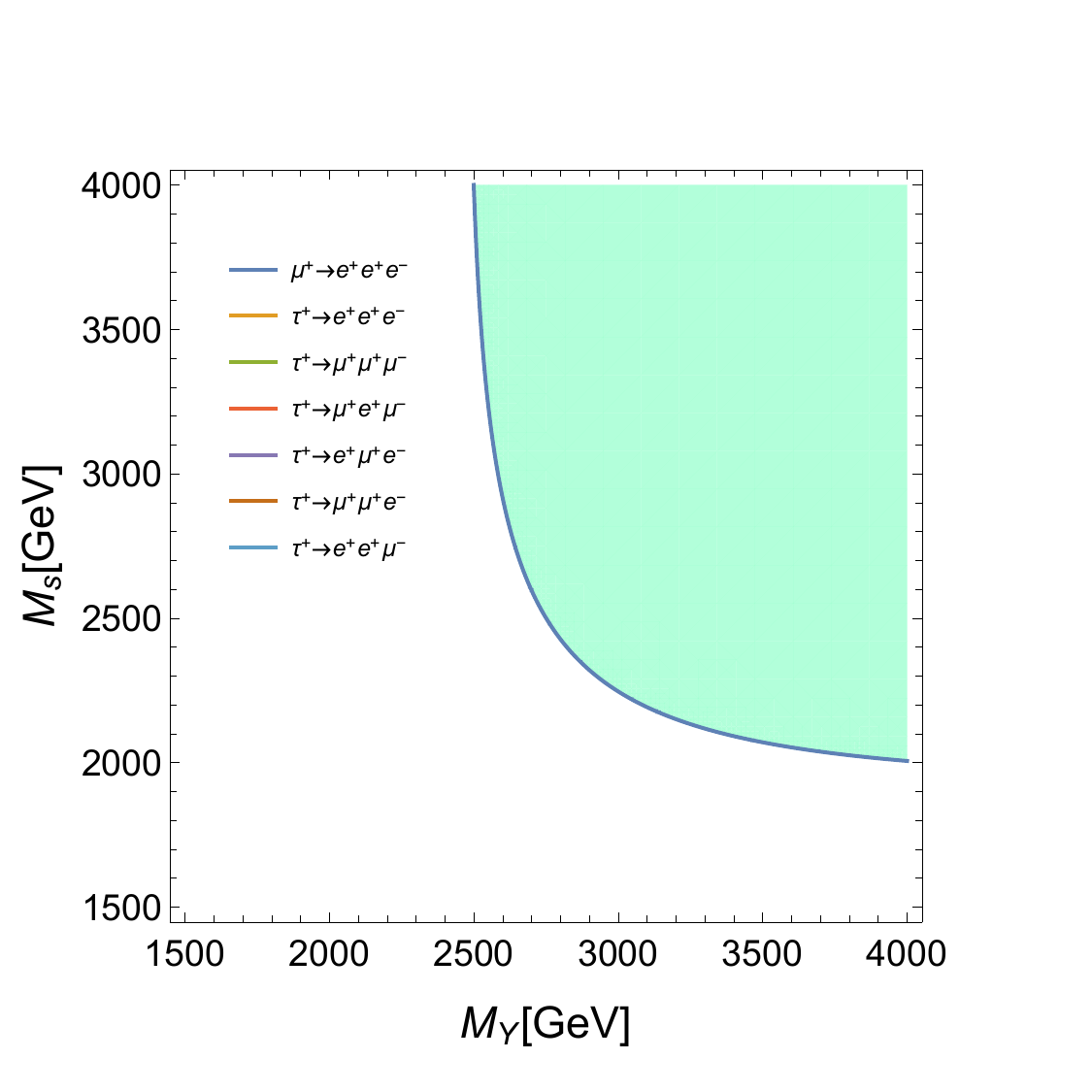}
		\caption{$|O_{Yij},O_{sij}|>10^{-2}$}
	\end{subfigure}
	\begin{subfigure}{0.5\linewidth}
	\centering
		\includegraphics[width=\linewidth,trim=0cm 0.4cm 0.8cm 1.2cm,,clip]{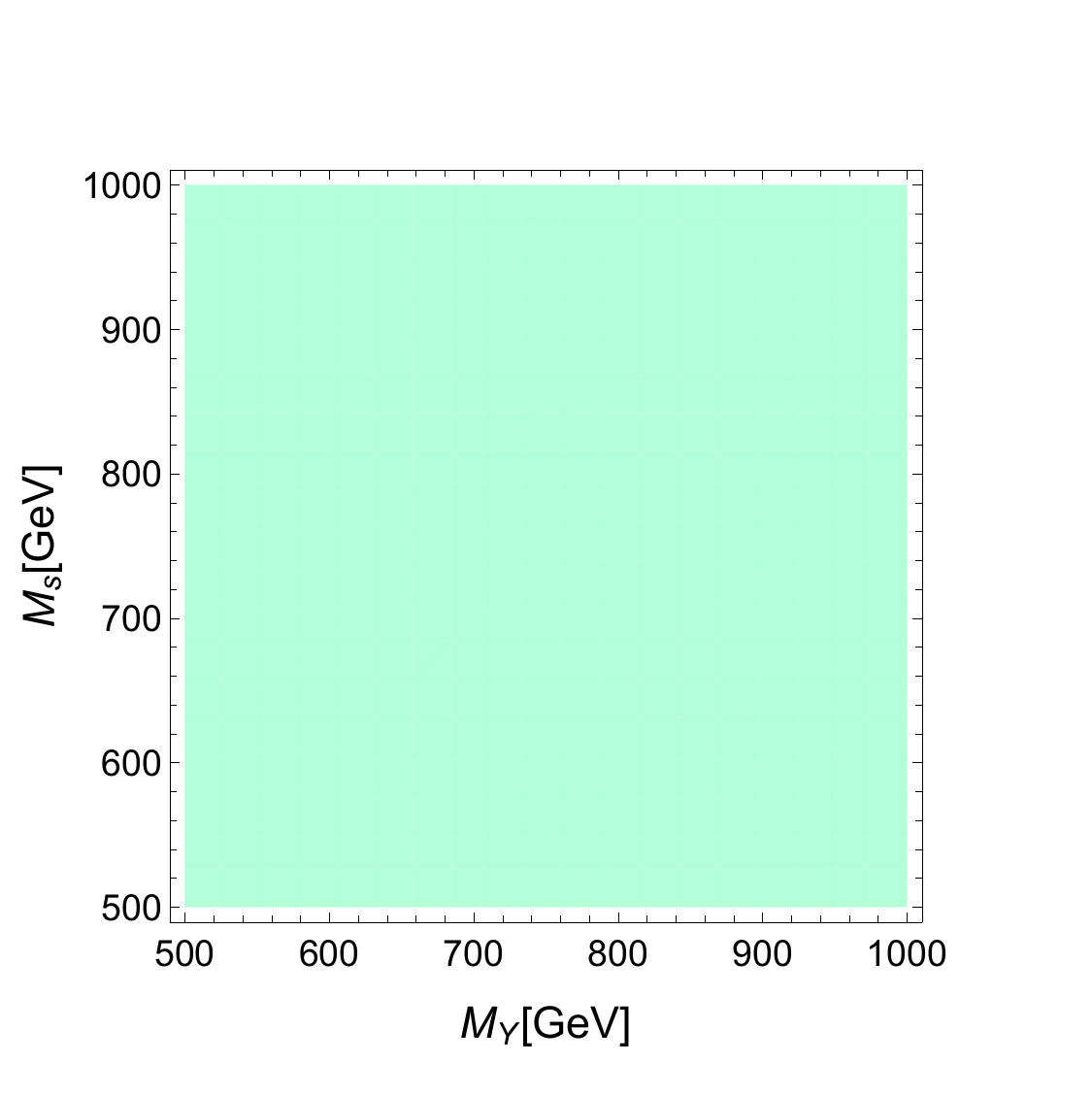}
		\caption{$|O_{Yij},O_{sij}|>10^{-3}$}
	\end{subfigure}}}
	\caption{Exclusion contours on the $M_Y \times M_s$ plane for bosons constrained to have similar masses.}
	\label{fig:contoursYs}
\end{figure}

The double scalar scenario is even less involved. The interference is, again, proportional to $m_e$, and there is no unitary mixing. Consequently, the structure of the solution is such that lower masses become allowed with diminishing couplings. We  enforce that the scalar masses are nearly degenerate, with which we see, from Figure~\ref{fig:contoursYs}, that couplings of the order $g_{\text{eff}} \sim O_{sij} \sim 10^{-2}$ allow for scalar masses of the order of \SI{2.5}{TeV}, while couplings as small as $10^{-3}$ are permissive of low masses. It is easy to notice that, in this case, since there is neither conditions tying different matrix elements together nor interference, the strongest bound, \textit{i.e.}, that of $\mu \to 3e$, is the only one that matters.

\section{Conclusions}
\label{sec:conc}

One important aspect in the search for new physics is knowing where to look, so that a common goal in BSM phenomenology is to impose constraints on exotic parameters, specially, lower bounds on masses. The high explorable energies and the excited stage of high amount of data collection reached by the LHC make it one ideal tool in this line of work. 

Specifically, it has been used to derive constraints on $M_U$, the mass of a doubly charged vector bilepton, rare feature of BSM models. The collective result of these LHC efforts and also of other kinds of searches, such as precision muonium-antimuonium conversion \cite{Gusso:2002cr}, is well described by the bound $M_U \gtrsim \SI{1}{TeV}$.

However, most works responsible for the bound above neglect the $V_U$ mixing of the $U\text{-}\ell \ell$ interaction \cite{Dutta:1994pd,Dion:1998pw,Barreto:2013paa,RamirezBarreto:2011av,RamirezBarreto:2008wq,Nepomuceno:2016jyr,Tully:1999yg,Meirose:2011cs,Corcella:2018eib,Corcella:2017dns,Frampton:2020pef,Coriano:2018khp}. This unitary mixing matrix is predicted by a skeptical analysis and definition of the interaction form and, in the 3-3-1 model, cannot be diagonal. This is not pragmatically problematic since the corresponding studies' objects are flavor conserving processes. It remained necessary to perform, for the vector bilepton, phenomenology useful within the alternative, larger sector of parameter space containing finite mixing.

This is the aim of this work, which shows that the simple bounds on branching ratios of 3-body lepton decays produce strong constraints on the bilepton mass in that case. To see this, it is enough to regard the pure $U^{\pm \pm}$ scenario, in which our study predicts $M_U> \SI{3200}{GeV}$ if the hierarchy within $V_U$ is of the order or lesser than $10^{3}$. It may then be visualized that the CLFV lepton decays bounds complement the LHC flavor-diagonal phenomenology, and is, furthermore, considerably more effective in the case of finite mixing if compared with Ref.~\cite{Barela:2019pmo}, which considers a CLFV LHC process, treating $V_U$ through a simplified construction.

The advantages of the purely leptonic processes are operationally manifest: although the 3-body phase space, relevant for the CLFV lepton decays, is substantially more computationally demanding than the 2-body one, the hadron physics needed for LHC phenomenology is an immensely heavier complication.

We don't primarily intend to achieve new specific mass bounds for the scalars: these particles interactions' are not governed by unitary mixing, and there are concrete (model-dependent) experimental bounds for the doubly-charged scalar \cite{ATLAS:2017xqs,ATLAS:2014kca}, and the neutral scalar is a well known and common particle, analogous to the Higgs boson, so that its phenomenology is well understood in most models where it is present \cite{ATLAS:2017eiz, CMS:2018rmh, ATLAS:2019nkf, ATLAS:2018bnv}. Nevertheless, we consider a pure scalar $Y-s$ scenario and what we find is that for low masses to be possible after enforcement of the CLFV bounds, the effective coupling must be of order of $10^{-3}$. For comparison, the corresponding SM $H\text{-}ee$ and $H\text{-}\tau \tau$ couplings are given by $(\mathcal{O})_e \sim 2.07 \times 10^{-6}$ and $(\mathcal{O})_\tau \sim 7.24 \times 10^{-3}$, indicating that it is certainly reasonable for an exotic flavor violating $C\!P$-even neutral scalar, generally associated with higher characteristic mass scales, to possess interactions parametrized by effective couplings of the order necessary for its mass to be possibly low.

The addition of the scalar bosons to our analysis is mainly intended to aid us understand the part that secondary, non-dominant, particles can play altering the naive (single particle) exclusion contours of dominant degrees of freedom, which, in the present context, occurs when it is considered together with the $U^{\pm \pm}$. We observe that the balance between the $U^{\pm \pm}$ and $s$ contributions occurs in a manner that, in the optimal interference region, the lower bound on the mass of the $U^{\pm \pm}$ is relieved by ~20\%. Although it could be argued that such phenomenon can only happen in small, fine-tuned regions of parameter space, this behavior can happen in a general multi-particle scenario, specially in ones where a subset of parameters is constrained by exterior phenomenological or theoretical input, like the fitting of well measured distinct masses or mixing parameters, such as, for instance, a PMNS-like matrix which, in a given model, is dependent on the lepton mixing matrices $V_{L,R}$.

\section*{ACKNOWLEDGEMENTS}
MB would like to thank CNPq for the financial support and is specially grateful to Vicente Pleitez and Rodolfo Capdevilla for several useful discussions. JMD thanks to the Conacyt program Investigadoras e Investigadores por M\'exico, project 1753.

\appendix

\section{Deriving the amplitudes}\label{app:amps}

To correctly derive amplitudes from Lagrangians with explicit charge conjugation can be troublesome. The issue appears because when these fields make up the interaction there is generally more than a way to contract a spinor chain with initial and final states, in which case simply writing the vertices with an explicit charge conjugation matrix is not by itself a well defined and unambiguous process. Below we show how to arrive at the amplitudes corresponding to the doubly-charged vector and scalar boson mediation, which suffer from this complication. We follow the algorithm and refer to the description of Refs.~\cite{Denner:1992vza,Denner:1992me}, but focus on the matter of dealing with Lagrangians with explicit charge conjugation, in the form as would naturally emerge from a renormalizable fundamental gauge theory.

We begin defining how to write down the spinor structure. Each spinor line in a diagram will come together with an Arbitrary Fermion Flow Arrow (AFFA) -- recall that the true fermion flow is not continuous in this type of graph. With reference to this arbitrarily drawn line and the true fermion flow arrow, the rules for external fermion lines are

\begin{figure}[H]
\label{fig:spinorsrules}
\centering
\includegraphics[width=0.7\linewidth]{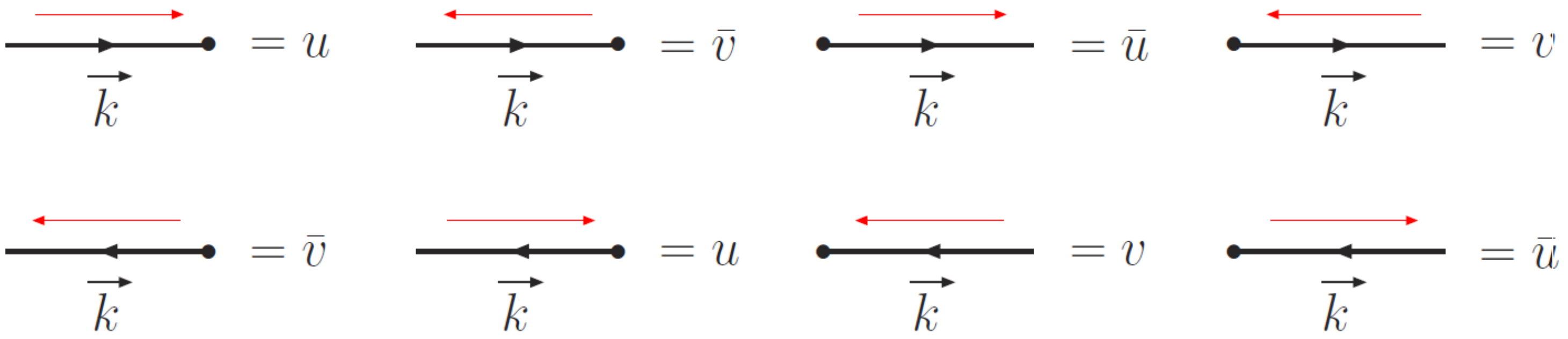}
\end{figure}

We also need to read the vertices off of the Lagrangians (\ref{eq:Uint}) and (\ref{eq:Yint2}). Considering always incoming bosons, we have a first set of vertices, corresponding to the case in which the AFFA ends on the heaviest fermion, for each charge of the boson:

\begin{equation}
\includegraphics[height=2.3cm,valign=c]{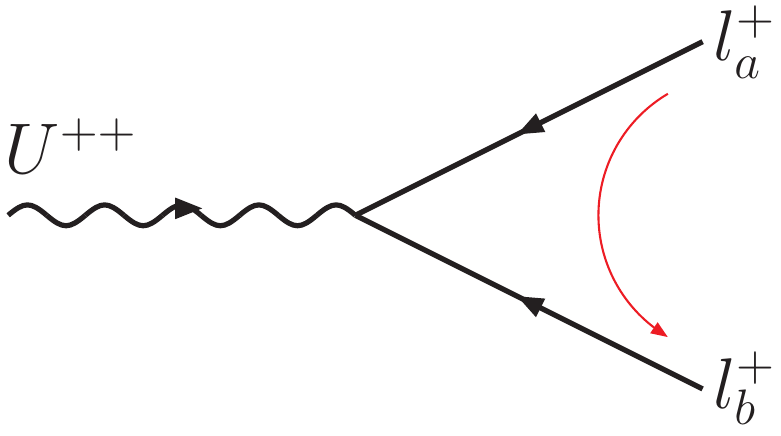} \equiv\Gamma^{U^{++}}_{ba}=\frac{g_{2L}}{\sqrt{2}}\gamma^\mu [P_L (V_U)_{ab}-P_R (V_U)_{ba}],
\end{equation}

\begin{equation}
\includegraphics[height=2.3cm,valign=c]{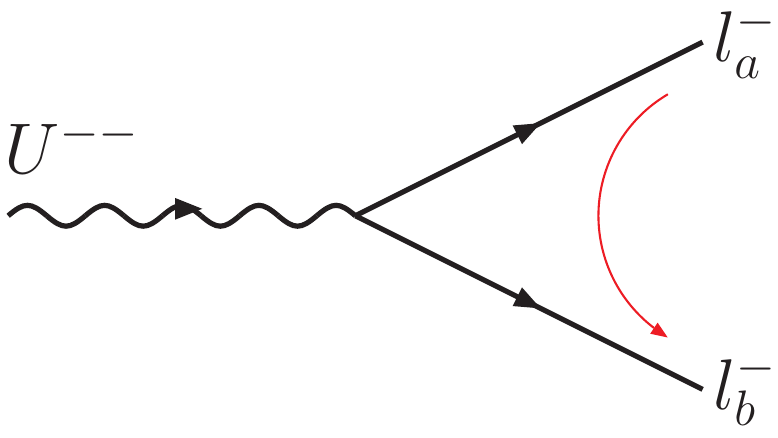} \equiv\Gamma^{U^{--}}_{ba}=\frac{g_{2L}}{\sqrt{2}}\gamma^\mu [P_L (V_U^\dagger)_{ab}-P_R (V_U^\dagger)_{ba}],
\end{equation}
The formulas above are valid even when $a=b$, which can be seen symmetrizing the corresponding part of Lagrangian (\ref{eq:Uint}) as $\bar{\ell_a^c}\gamma^\mu P_L  \ell_a=\frac{1}{2}\left[\bar{\ell_a^c}\gamma^\mu P_L  \ell_a - \bar{\ell_a^c}\gamma^\mu P_R  \ell_a\right]$. \footnote{Notice again that the vector part of this interaction dies out.}The remaining relative factor of $1/2$ is compensated in the rule by a factor of $2$ due to the identical particles. These vertices are called \textit{regular}.

The seemingly innocuous choice of leaving the heaviest fermion on the right in the Lagrangians made in Section \ref{sec:effint} is what amounts to defining the above vertices as the regular ones.

The second set of vertices for the $U^{\pm \pm}$ is obtained conjugating the original vertex by the charge conjugation matrix like $\Gamma^\prime=C \Gamma C^{-1}$ --  this recipe comes directly by transposition and manipulation of the reference spinor chain. We have that $C\gamma^\mu [P_L (V_U)_{ab}-P_R (V_U)_{ba}]C^{-1}=\gamma^\mu [P_L (V_U)_{ba}-P_R (V_U)_{ab}]$, so that the new vertex rule is $\Gamma^\prime_{ab}=\Gamma_{ba}$. We write the reversed vertices for completeness (recall that $\ell_b$ is the heaviest of the 2 leptons)

\begin{equation}
\includegraphics[height=2.3cm,valign=c]{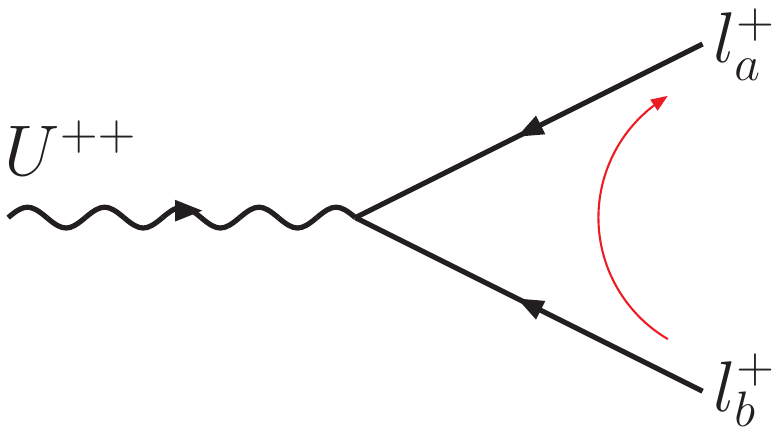}=\Gamma^{\prime U^{++}}_{ab}=i\frac{g_{2L}}{\sqrt{2}}\gamma^\mu [P_L (V_U)_{ba}-P_R (V_U)_{ab}],
\end{equation}

\begin{equation}
\includegraphics[height=2.3cm,valign=c]{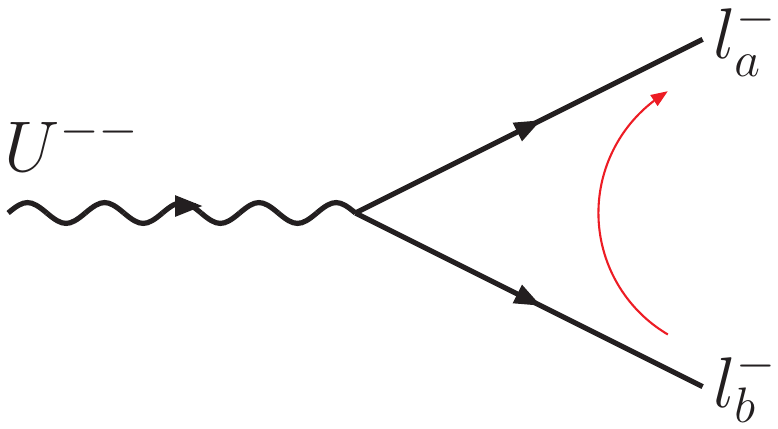}=\Gamma^{\prime U^{++}}_{ab}=i\frac{g_{2L}}{\sqrt{2}}\gamma^\mu [P_L (V_U^\dagger)_{ba}-P_R (V_U^\dagger)_{ab}].
\end{equation}

As an example up to this point, we write the rule corresponding to the two different choices of AFFA for the subdiagrams (and not vertices representations) below

\begin{equation}
\includegraphics[height=2.3cm,valign=c]{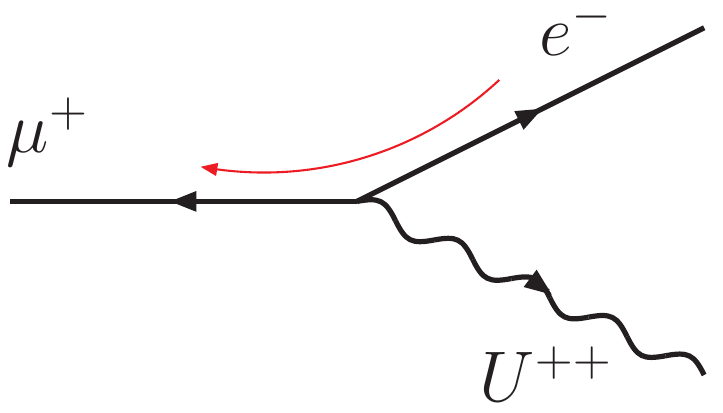}=\bar{v}_\mu \Gamma^{U^{--}}_{\mu e} u_e,
\end{equation}

\begin{equation}
\includegraphics[height=2.3cm,valign=c]{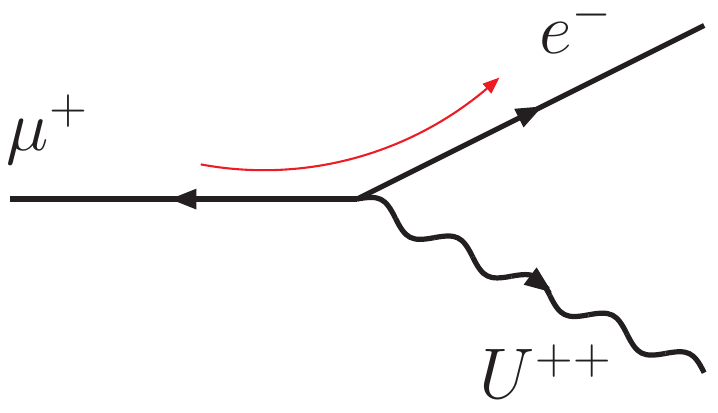}=\bar{v}_e \Gamma^{\prime U^{--}}_{e \mu} u_\mu.
\end{equation}

The vertices for the doubly-charged scalar are shown below

\begin{equation}
\includegraphics[height=2.3cm,valign=c]{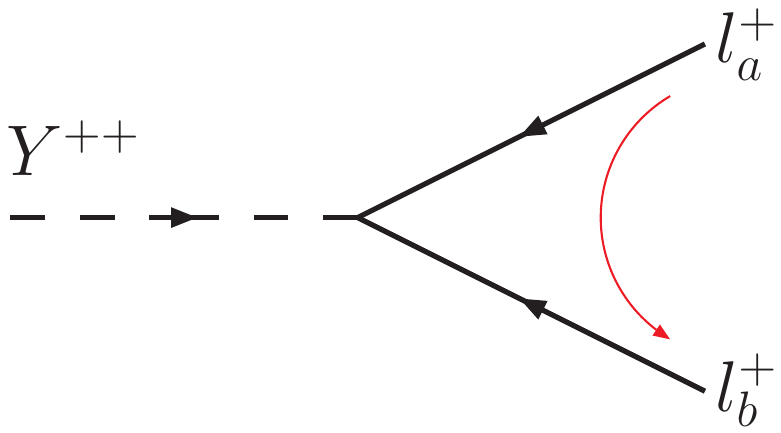}=\Gamma^{Y^{++}}_{ba}=-i g_{YL} \left[(\mathcal{O}_Y)_{ab}+(\mathcal{O}_Y)_{ba}\right] P_L,
\end{equation}

\begin{equation}
\includegraphics[height=2.3cm,valign=c]{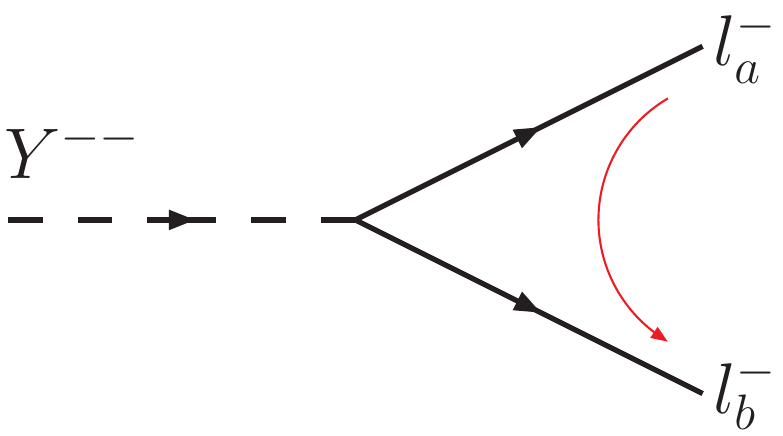}=\Gamma^{Y^{--}}_{ba}=-i g_{YL} \left[(\mathcal{O}_Y^\dagger)_{ab}+(\mathcal{O}_Y^\dagger)_{ba}\right] P_R,
\end{equation}

\begin{equation}
\includegraphics[height=2.3cm,valign=c]{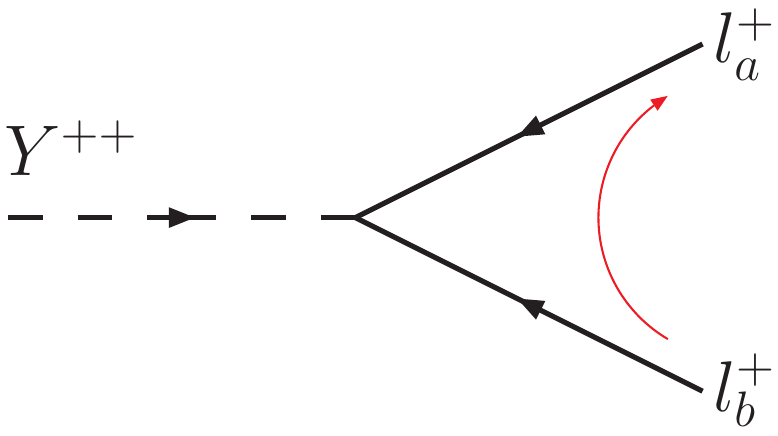}=\Gamma^{\prime Y^{++}}_{ab}=-i g_{YL} \left[(\mathcal{O}_Y)_{ba}+(\mathcal{O}_Y)_{ab}\right] P_L,
\end{equation}

\begin{equation}
\includegraphics[height=2.3cm,valign=c]{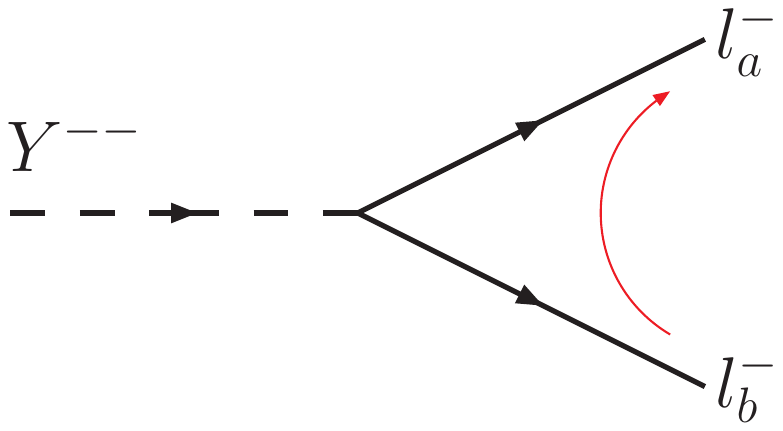}=\Gamma^{\prime Y^{--}}_{ab}=-i g_{YL} \left[(\mathcal{O}_Y^\dagger)_{ba}+(\mathcal{O}_Y^\dagger)_{ab}\right] P_R.
\end{equation}
We emphasize one last time that what defines if a vertex is regular or reversed is the direction of the AFFA with respect to fermion generation -- which, in turn, is a consequence of the conventional form of the Lagrangian.

Knowing the vertices and how to write the exotic spinor chains, the missing ingredient is the ability to find the relative sign between diagrams. This is the greatest reason for the necessity of an algorithm that substitutes the mere explicit use of the charge conjugation matrix. Within the algorithm, to find the relative signs amounts to simply comparing particle "order" -- more precisely, the order in which spinors appear in the chain -- with respect to the AFFA and identifying the order of the relating permutation.

Refer to our real diagrams of Fig.~\ref{fig:diagrams}. The particle orders are (we label different particles by the momenta)

\begin{equation}
\begin{split}
R(\mathcal{M}_U)&=(p,k_3,k_1,k_2), \\
R(\mathcal{M}_Y)&=(p,k_3,k_1,k_2), \\
R(\mathcal{M}_{s1})&=(p,k_1,k_3,k_2), \\
R(\mathcal{M}_{s2})&=(p,k_2,k_3,k_1).
\end{split}
\end{equation}
Taking $R(\mathcal{M}_U)$ as the referential, we identify that the only ordered set related to it by an odd permutation is $R(\mathcal{M}_{s1})$, so that $i\mathcal{M}_{s1}$ comes attached to an extra minus sign.

This concludes a sufficient description of how our amplitudes can be obtained from the given Lagrangians without having to appeal to an explicit analysis of the possible wick contractions involved in the correlator.

\newpage
\section{Solutions}
\label{app:sol}

\begin{table}[H]
\centering
\caption{Solutions of the pure $U$ scenario, corresponding to the plots of Fig.~\ref{fig:contoursU}.}\label{Tab:solU}
\begin{tabular}{c@{\hskip 1.5em}c@{\hskip 38pt}c@{\hskip 1.5em}c@{\hskip 38pt}r@{\hskip 1.5em}r}
\toprule
\multicolumn{4}{c}{Pure $U$: $\;|V_{Uij}|>10^{-3}$} \\ \midrule
$M_U$      &   $3380$       &  $\psi$     &   $1.48108$ \\
$\phi$     &   $3.03983$    &  $\theta$   &   $2.66989$ \\ \toprule
\multicolumn{4}{c}{Pure $U$: $\;|V_{Uij}|>10^{-4}$} \\ \midrule
$M_U$      &   $1100$       &  $\psi$     &   $0.78535$ \\
$\phi$     &   $5.49774$    &  $\theta$   &   $0.00014$ \\ \toprule
\multicolumn{4}{c}{Pure $U$: $\;|V_{Uij}|>10^{-5}$} \\ \midrule
$M_U$      &   $500$       &  $\psi$     &   $0.72066$ \\
$\phi$     &   $0.72067$   &  $\theta$   &   $3.13801$ \\ \toprule
\multicolumn{4}{c}{Pure $U$: $\;|V_{Uij}|>10^{-5}$, $V_{U11}=0.85$} \\ \midrule
$M_U$      &   $6830$       &  $\psi$     &   $4.74012$ \\
$\phi$     &   $4.68301$    &  $\theta$   &   $0.55504$ \\ \toprule
\end{tabular}
\end{table}

\begin{table}[H]
\centering
\caption{Solutions of the $U-s$ scenario, respective of the plots shown in Fig.~\ref{fig:contoursUs}}\label{Tab:solUs}
\begin{tabular}{c@{\hskip 1.5em}c@{\hskip 38pt}c@{\hskip 1.5em}c@{\hskip 38pt}r@{\hskip 1.5em}r}
\toprule
\multicolumn{6}{c}{$U-s$: $\;|V_{Uij},\mathcal{O}_{sij}|>10^{-3}$} \\ \midrule
$M_U$      &   $2650$       &  $\mathcal{O}_{s11}$  &   $3.11564 \times 10^{-3}$ & $\mathcal{O}_{s22}$   &  $1.43763 \times 10^{-3}$  \\
$M_s$      &   $500$        &  $\mathcal{O}_{s12}$  &   $3.01898 \times 10^{-3}$ & $\mathcal{O}_{s23}$   &  $4.06257 \times 10^{-2}$  \\
$\phi$     &   $6.26303$    &  $\mathcal{O}_{s13}$  &   $4.20596 \times 10^{-2}$ & $\mathcal{O}_{s31}$   &  $-1.32129 \times 10^{-1}$ \\
$\psi$     &   $1.55218$    &  $\mathcal{O}_{s21}$  &   $3.19852 \times 10^{-3}$ & $\mathcal{O}_{s32}$   &  $1.51094 \times 10^{-1}$  \\
$\theta$   &   $2.90919$    &                       &  \\ \toprule
\multicolumn{6}{c}{$U-s$: $\;|V_{Uij},\mathcal{O}_{sij}|>10^{-4}$} \\ \midrule
$M_U$      &   $840$       &  $\mathcal{O}_{s11}$  &   $\hphantom{-}2.57667 \times 10^{-3}$ & $\mathcal{O}_{s22}$   &  $-2.14209 \times 10^{-3}$  \\
$M_s$      &   $500$       &  $\mathcal{O}_{s12}$  &   $-3.29400 \times 10^{-3}$ & $\mathcal{O}_{s23}$  &  $-6.05650 \times 10^{-2}$  \\
$\phi$     &   $1.45901$   &  $\mathcal{O}_{s13}$  &   $\hphantom{-}4.05147 \times 10^{-1}$ & $\mathcal{O}_{s31}$   &  $-4.81308 \times 10^{-2}$ \\
$\psi$     &   $1.45911$   &  $\mathcal{O}_{s21}$  &   $-3.34363 \times 10^{-3}$ & $\mathcal{O}_{s32}$  &  $-1.44602 \times 10^{-1}$  \\
$\theta$   &   $3.13998$    &                       &  \\ \toprule
\multicolumn{6}{c}{$U-s$: $\;|V_{Uij},\mathcal{O}_{sij}|>10^{-5}$} \\ \midrule
$M_U$      &   $<500$       &  $\mathcal{O}_{s11}$  &   $1.00000 \times 10^{-5}$ & $\mathcal{O}_{s22}$   &  $1.00000 \times 10^{-5}$  \\
$M_s$      &   $<500$       &  $\mathcal{O}_{s12}$  &   $1.00000 \times 10^{-5}$ & $\mathcal{O}_{s23}$  &  $1.00000 \times 10^{-5}$  \\
$\phi$     &   $0.72067$   &  $\mathcal{O}_{s13}$  &   $1.00000 \times 10^{-5}$ & $\mathcal{O}_{s31}$   &  $1.00000 \times 10^{-5}$ \\
$\psi$     &   $0.72066$   &  $\mathcal{O}_{s21}$  &   $1.00000 \times 10^{-5}$ & $\mathcal{O}_{s32}$  &  $1.00000 \times 10^{-5}$  \\
$\theta$   &   $3.13801$    &                       &  \\ \toprule
\multicolumn{6}{c}{$U-s$: $\;|V_{Uij},\mathcal{O}_{sij}|>10^{-5}$, $\mathcal{O}_{s11}=\mathcal{O}_{s22}=1$} \\ \midrule
$M_U$      &   $1800$      &  $\mathcal{O}_{s11}$  &  $1.00000$                 & $\mathcal{O}_{s22}$  &  \multicolumn{1}{c}{$1.00000$}  \\
$M_s$      &   $580$       &  $\mathcal{O}_{s12}$  &  $-1.12685 \times 10^{-5}$ & $\mathcal{O}_{s23}$  &  $-8.60549 \times 10^{-4}$  \\
$\phi$     &   $0.00020$   &  $\mathcal{O}_{s13}$  &  $-1.19611 \times 10^{-3}$ & $\mathcal{O}_{s31}$  &  $1.83141 \times 10^{-4}$ \\
$\psi$     &   $0.00024$   &  $\mathcal{O}_{s21}$  &  $-1.13022 \times 10^{-5}$ & $\mathcal{O}_{s32}$  &  $2.55491 \times 10^{-3}$  \\
$\theta$   &   $3.09077$   &                       &  \\ \toprule
\end{tabular}
\end{table}

\end{document}